\PassOptionsToPackage{bookmarksnumbered,unicode}{hyperref}
\documentclass[sigconf]{acmart}

\AtBeginDocument{%
  \providecommand\BibTeX{{%
    \normalfont B\kern-0.5em{\scshape i\kern-0.25em b}\kern-0.8em\TeX}}}

\copyrightyear{2026}
\acmYear{2026}
\setcopyright{cc}
\setcctype{by-nc-nd}

\acmConference[UIST '26]{The 39th Annual ACM Symposium on User Interface Software and Technology}{November 02--05, 2026}{Detroit, MI, USA}
\acmBooktitle{The 39th Annual ACM Symposium on User Interface Software and Technology (UIST '26), November 02--05, 2026, Detroit, MI, USA}
\acmDOI{10.1145/3830398.3830681}
\acmISBN{979-8-4007-2856-3/2026/11}

\usepackage{amsmath}
\usepackage{physics}
\usepackage{array}

\usepackage{enumitem}
\usepackage{xspace}

\usepackage{xcolor}
\definecolor{myblue}{rgb}{0.216,0.494,0.965}

\usepackage{todonotes}

\newcommand{\revisionchanges}[1]{#1}

\def\splatxr{\textsc{SplatXR}\xspace}

\begin{document}

\title[SplatXR]{SplatXR: A Toolkit for Multi-Layered Extended Reality via On-Headset 3D Gaussian Splatting}

\author{Cyrus Vachha}
\affiliation{
  \institution{Princeton University}
  \city{Princeton, NJ}
  \country{USA}}
\email{cvachha@princeton.edu}

\author{Mohamed Kari}
\affiliation{
  \institution{Princeton University}
  \city{Princeton, NJ}
  \country{USA}}
\email{mokari@princeton.edu}

\author{Parastoo Abtahi}
\affiliation{
  \institution{Princeton University}
  \city{Princeton, NJ}
  \country{USA}}
\email{parastoo@princeton.edu}

\renewcommand{\shortauthors}{Vachha et al.}

\begin{abstract}
    Recent advances in Gaussian splatting, 3D foundation models, and video-passthrough mixed reality offer unprecedented fidelity for capturing and blending physical and virtual worlds. However, established XR methods rely on polygon-based pipelines that treat physical and digital environments as separate domains, making seamless integration across recent novel representations difficult. Because splats lack explicit surface geometry, they impede standard graphics techniques necessary for cross-reality interactions, such as shadow or collision handling. Furthermore, while 3D foundation models can reconstruct physical objects on-the-fly into digital splats, \revisionchanges{current XR systems fall short of leveraging this capability for user experiences that} transition seamlessly between physicality and virtuality. \revisionchanges{To address this, we organize XR as a multi-layered reality of interacting primitives, letting pixels and objects move across physical and digital boundaries. We contribute \splatxr, a Unity toolkit that operationalizes this structure as building blocks for interactive, immersive apps leveraging 3D Gaussian splatting:} a standalone renderer, splat-to-mesh reconstruction, splat physics, splat blending, and offloaded splat object reconstruction via on-headset capture. \revisionchanges{\splatxr unites vision and graphics techniques to make splat-based XR development broadly available,} and we validate it through demonstrative applications and a small developer study. \looseness=-1
\end{abstract}



\begin{teaserfigure}
    \includegraphics[width=\linewidth]{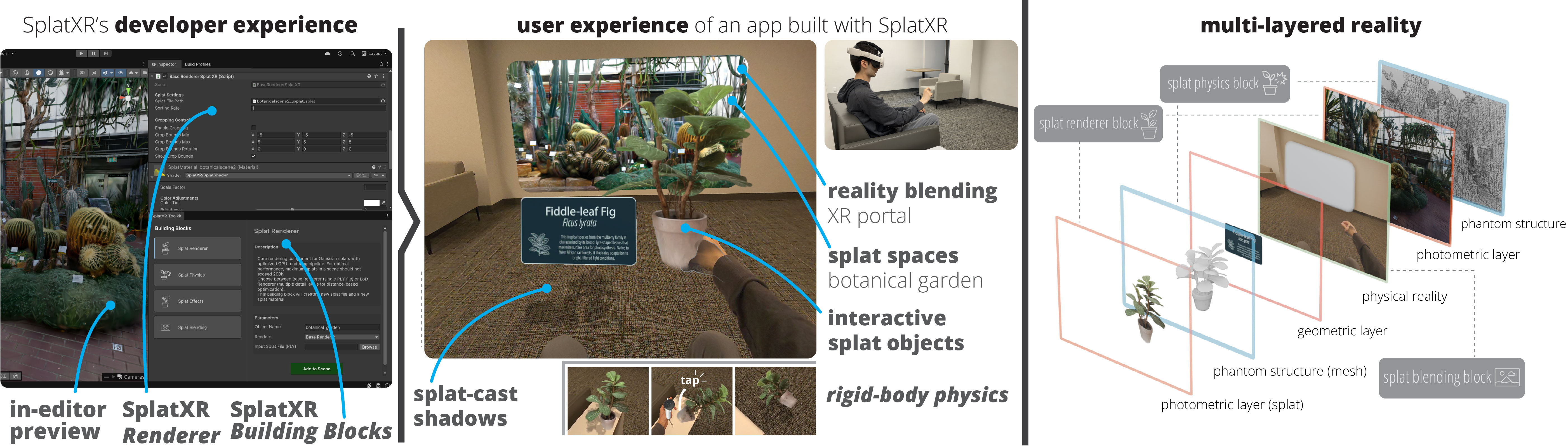}
    \vspace{-7mm}
    \Description{Figure 1 has three horizontal subpanels. The first, captioned "SplatXR's developer experience," is a screenshot of the Unity editor. The game view on the left shows a botanical garden. The Unity inspector, top right, is set to a game object displaying a script named "Splat Renderer." Below it is a panel tabbed "SplatXR Toolkit" with two columns: on the left, four stacked buttons: "Splat Renderer," "Splat Physics," "Splat Effects," "Splat Blending"; on the right, a description of the Splat Renderer, three fields ("Object Name," "Renderer," "Input Splat File (PLY)"), and an "Add to Scene" button. Annotation lines label the game view "in-editor preview," the inspector script "SplatXR Renderer," and the toolkit panel "SplatXR Building Blocks." An arrow points to the next subfigure. The second, "user experience of an app built with SplatXR," is a mixed-reality headset screenshot with smaller images around it. A simple room with a white wall and carpet is visible in passthrough. A virtual portal on the wall shows the same botanical garden from a new perspective. The user's right arm extends toward the portal holding a photorealistic virtual potted plant, labeled "Fiddle-leaf Fig, Ficus Lyrata." A plant-shaped shadow falls on the carpet. Annotations label the portal "reality blending: XR portal" and "splat spaces: botanical garden," and the plant and its shadow "splat-cast shadows" and "interactive splat objects." A side image shows a user in a Meta Quest 3 holding the virtual plant. Three thumbnails, captioned "rigid-body physics: splat against world," show the plant on a table, a hand tapping it (marked "TAP"), and the plant fallen on the floor. The third, "multi-layered reality," shows six separated isometric layers, front to back: plant splat ("photometric layer (splat)"), blue-bordered plant mesh ("phantom structure (mesh)"), an info panel ("geometric layer"), the room with an outstretched hand ("physical reality"), the botanical garden ("photometric layer"), and a garden mesh ("phantom structure"). Labeled boxes link the splat rendering block to layers 1 and 5, the splat physics block to layers 2 and 6, and the splat blending block to layer 4.}
    \caption{
      \emph{Left:} A developer composes an XR app using our \splatxr toolkit in Unity.
      \emph{Mid:} The resulting app can be deployed to a standalone Meta Quest 3 or Apple Vision Pro headset, featuring several toolkit capabilities for a rich, interactive user experience.
      \emph{Right:} The toolkit leverages a set of custom \splatxr representations.
    }
    \label{fig:teaser}
\end{teaserfigure}

\maketitle

\section{Introduction}

In recent years, 
three fundamentally new ways of \emph{reconstructing, representing, and rendering a scene} have emerged. 
First, vision foundation models such as SAM 3 \cite{carion2025sam3segmentconcepts} and SAM 3D \cite{sam3dteam2025sam3d3dfyimages} have made it possible to gain a detailed understanding of 3D scenes and their semantics.
Second, differentiable rendering techniques such as neural radiance fields \cite{mildenhall2020nerf} and 3D Gaussian splatting (3DGS)~\cite{kerbl3Dgaussians} have enabled capturing complex 3D scenes with photorealism. 
Third, splat- or video-based world modeling (e.g., World Labs~\cite{worldlabs}) has the ability to simulate visual 3D worlds directly from the physical world, enabling the generation of interactive 3D environments from minimal visual input.
All of these novel techniques enable the creation of 3D worlds at unprecedented fidelity grounded in physical reality by leveraging implicit representations.\looseness=-1

At the same time, a new generation of video-passthrough head-worn displays, such as the Meta Quest 3, Apple Vision Pro, and Samsung Galaxy XR, has pushed the limits of \emph{displaying a scene} and the achievable visual integration of physical-digital scene compositions. 
These extended reality (XR) devices leverage perspective reprojection and high-resolution displays to obtain scenes that seamlessly blend virtual elements with physical reality.
However, despite the technical innovation underlying these new XR displays, their applications are still fundamentally based on and designed for displaying explicit polygon mesh representations, used for decades in computer graphics, as they provide visual information for rendering and geometric information for collisions, shadows, and occlusions.\looseness=-1


%
This mismatch between the explicit representations in traditional polygon meshes and the implicit representations driving today's computer vision and graphics innovations has hindered the adoption of the latter for interactive purposes in XR.
\revisionchanges{With this work, we introduce a toolkit that is organized around a multi-layered composition} that accounts for the complementary strengths and shortcomings of explicit and implicit scene representations, all while ensuring ease-of-use for XR application developers.
In particular, we focus on 3DGS as it is at the core of many graphics innovations, also serving as a common interface to vision foundation models and world models (e.g., SAM 3D \cite{sam3dteam2025sam3d3dfyimages} can export its reconstruction as a splat, and Marble by World Labs \cite{worldlabs} can also export its generation as a splat). 
We organize these physical and digital layers into a \revisionchanges{manageable compositional structure}, given that perceived reality is a composite of various structural, visual, and semantic components. By treating reality as a set of distinct but interacting layers across different representations (physical reality, Gaussian splats, environmental meshes, scene semantics), our framework allows developers to design cross-reality interactions where pixels and objects can move between these layers.
To operationalize this, we lay out several example configurations that recur across different layers of reality. 
For instance, structural layers can influence visual layers through \revisionchanges{proxy} phantom meshes, i.e., invisible geometries reconstructed from the physical environment which let virtual splats cast shadows and appear to collide with real-world surfaces. 
We showcase configurations for \emph{semantic linking} across layers, demonstrating how inputs, such as hand tracking, can be used to anchor and manipulate digital splat objects. 
This provides developers with the mental model and technical foundation necessary to build experiences that bidirectionally interact with the real environment and captured or generated 3D scenes. 


\revisionchanges{We contribute \splatxr, an end-to-end Unity toolkit for building XR applications with Gaussian splats on standalone, untethered headsets, integrating real-time splat rendering, physics, and developer-facing abstractions.
\splatxr thus facilitates an integrated developer workflow that brings together and adapts computer vision and graphics techniques in one toolkit so that building XR applications with Gaussian splatting becomes available to the wider XR community, from research to development.
Through building blocks, \splatxr automatically handles proxy geometry computation, Gaussian splat materials, and render queue ordering, maintaining visual and physical consistency across splats, meshes, and proxy geometry without additional developer input.
Unlike prior splatting work confined to VR, \splatxr also enables AR by simultaneously running wire- and server-free for unbounded outdoor use and providing AR-specific support, including passthrough compositing, shader-based alpha blending, color correction, and occlusion from reconstructed environment geometry.
The toolkit also enables interactive reconstruction of objects directly from the headset, offloading compute to a desktop GPU running generative methods, so developers can integrate realistic digital object twins into applications to bridge physical and digital spaces.}
We showcase applications built with \splatxr demonstrating how different elements are linked semantically between layers and interact with one another. 
\revisionchanges{A technical evaluation of the toolkit features and renderer shows the efficacy of the implementation, and} a small preliminary developer evaluation (n=3) provides initial insights into the toolkit's usability.
\revisionchanges{\splatxr and our technical evaluation setup are open sourced and available online\footnote{\url{https://splatxr.github.io}}.}\looseness=-1


\section{Related Work}

\splatxr is a toolkit that enables developers to leverage Gaussian splats of captured physical objects and spaces for interactive XR applications.
Its contributions therefore relate to cross-reality interaction, capture-based XR experiences, neural scene representations in XR, XR toolkits, and the emerging space of Gaussian splatting software tools and libraries. \looseness=-1

\subsection{Cross-Reality Interactions}
\revisionchanges{Cross-reality interactions have been conceptualized in a variety of ways to support different research or development purposes.}
Cools et al.~\cite{cools2025display} examine the compositional structure of cross-reality interactions for design patterns.
Auda et al.~\cite{auda2023scoping} survey cross-reality systems along the reality-virtuality continuum.
Blending Spaces~\cite{cools2022blending} transitions between physical and virtual layers on an object level, and RealityCheck~\cite{hartmann2019realitycheck} does so on a pixel level. GradualReality~\cite{seo2024gradualreality} further incorporates interaction state awareness.
All of these works assume---even if implicitly---that the virtual world consists of polygon meshes and thus forms a single coherent layer. 
Therefore, they do not account for the challenges that arise in using Gaussian \revisionchanges{splatting for interactive and situated applications, e.g., ensuring lighting coherence or supporting physics simulations between explicitly and implicitly represented content}.
We instead implement a Unity toolkit that accounts for both.

\subsection{Capture-Based XR Experiences}
Several smartphone-based apps scan physical objects into 3D representations for downstream XR applications.
While meshes have predominated, recent products from Luma~\cite{lumalabs}, Scaniverse~\cite{scaniverse}, and Polycam~\cite{polycam} export splats; \splatxr expects developers to provide a splat created with such an app.\looseness=-1

Captured 3D models underpin a multitude of XR experiences: remote robot operation in VRoxy~\cite{sakashita2023vroxy}, environmental design prototyping in PointShopAR~\cite{wang2023pointshopar}, redirected walking in SwitchAR~\cite{wombacher2025switchar}, collaboration in VirtualNexus~\cite{huang2024virtualnexus}, interface personalization in MineXR~\cite{cho2024minexr}, educational experiences in Museum in the Classroom~\cite{ly2025museum}, memorable item interaction in InteRecon~\cite{li2025interecon}, and animating the physical environment in Scene Responsiveness~\cite{kari2023scene}.
SplatOverflow~\cite{10.1145/3706598.3714129} anchors annotations to splats for asynchronous hardware troubleshooting, and Dreamcrafter~\cite{vachha2025dreamcrafter} edits and creates VR scenes from captured and AI-generated splats.
Others virtually represent physical objects with hand-designed 3D models~\cite{pointecker2024real,fender2022causality}.
\revisionchanges{Beyond specific experiences, XR-Objects~\cite{xrobjects} contributes MLLM-driven interactions with everyday objects, and Programmable Reality~\cite{programmablereality} outlines a vision for programmable AR. Foundational depth reconstruction and tracking methods such as KinectFusion~\cite{kinectfusion} reconstruct the environment geometry that can be used for occlusion and physics.}
These works demonstrate modular components or potential use cases that motivate \splatxr's goal to simplify splat integration into XR development \revisionchanges{and to provide toolkit features for recurring problems such as object import, rendering, and interaction.}\looseness=-1

\subsection{Implicit Scene Representations in XR}
The use of splats and other implicit scene representations in XR has ranged from rendering to interaction.
VRSplat~\cite{tu2025vrsplat} and VR-Pipe~\cite{lee2025vrpipestreamlininghardwaregraphics} tackle the challenges of rendering splats under headset constraints, MetalSplatter~\cite{metalsplatter} enables splat rendering on Apple Vision Pro, and Snap's Lens Studio~\cite{lensstudio} does the same on Snap Spectacles.
\revisionchanges{VR-GS~\cite{jiang2024vr} deforms otherwise static splats for interaction purposes, and GS-Verse~\cite{pechko2025gsversemeshbasedgaussiansplatting} similarly enables single-object soft-body deformation by building on the mesh-based GaMeS representation~\cite{2024gamesmesh}. However, both are tethered and VR-only, targeting neither passthrough AR nor an easily importable toolkit. Reality Promises~\cite{realitypromises} renders untethered scene and object splats, but does not offer a modular toolkit or mesh-splat interactions.}
Li et al.~\cite{li2023interacting} render neural radiance fields (NeRFs) on a headset with simple transform manipulations, Instant-NGP~\cite{mueller2022instant} allows VR viewing of NeRFs and cleaning of floaters, and Splatshop~\cite{schutz12025splatshop}, VR-Doh~\cite{luo2025vrdoh}, and GaussianShopVR~\cite{shen2025GaussianShopVR} facilitate editing splats in VR.
GaussianNexus~\cite{gaussiannexus}, CoCreateAR~\cite{Numan_2025},
Reality-Linked Gaussian-Splatted Proxies~\cite{Yangetal2026},
Thing2Reality~\cite{10.1145/3672539.3686740}, and SharedNeRF~\cite{sakashita2024sharednerf} implicitly represent objects of interest across spatially distributed users for telepresence and remote collaboration. 
\revisionchanges{MiXR~\cite{mixr} similarly applies scanned-object interactions to a specific domain (generative-aided design) and, like GaussianNexus, targets a particular use case rather than a general-purpose XR toolkit.} 
Additional applications of splatting and NeRFs in XR are discussed by Qiu et al.~\cite{qiu2024advancingextendedreality3d} and Li et al.~\cite{li2025radiancefieldsxrsurvey}, who survey various use cases for radiance fields.\looseness=-1

Like prior work, \splatxr builds on several existing splatting techniques for on-headset rendering, 
\revisionchanges{and, to this end, unifies and packages established algorithms into a single, coherent XR toolkit for developers, adapting techniques as needed for rendering splats in an HMD.
Table~\ref{tab:comparison} provides a summary of existing systems.
\splatxr differs by (1) lowering the barrier to splat-based XR development by uniting computer vision and graphics techniques, (2) enabling AR-specific support, unlike prior work, which largely remains VR-only, and (3) providing reusable components for recurring XR tasks.}\looseness=-1

\begin{table}[t]
\centering
\small
\setlength{\tabcolsep}{4pt}
\caption{\revisionchanges{Splatting and XR systems compared to \splatxr. \emph{AR} denotes physical-digital compositing on headsets. \emph{Geom.} denotes geometry awareness for shadows, occlusions, and collisions. \emph{Splat} denotes rendering of 3DGS. \emph{Toolkit} denotes the availability of a package that can be simply imported into an existing project}.}
\label{tab:comparison}
\begin{tabular}{lccccc}
\toprule
& Untethered & AR & Geom. & Splat & Toolkit \\
\midrule
UnityGaussianSplatting~\cite{unitygs} & -- & -- & -- & \checkmark& --\\
Ninjamode Unity GS ~\cite{kleinbeck_multi-layer_2025ninjamode}& \checkmark& -- & -- & \checkmark& --\\
GS-Verse~\cite{pechko2025gsversemeshbasedgaussiansplatting} & -- & -- & -- & \checkmark & -- \\
VR-GS~\cite{jiang2024vr} & -- & -- & -- & \checkmark & -- \\
GaussianNexus~\cite{gaussiannexus} & -- & \checkmark & -- & \checkmark& -- \\
MiXR~\cite{mixr} & \checkmark& \checkmark& \checkmark& --& -- \\
Reality Promises~\cite{realitypromises} & \checkmark & \checkmark & --& \checkmark& -- \\
\midrule
\textbf{\splatxr (ours)} & \checkmark & \checkmark & \checkmark & \checkmark & \checkmark \\
\bottomrule
\end{tabular}
\end{table}

\subsection{XR Developer Toolkits}
Toolkits constitute an integral part of HCI research \cite{ledo2018evaluation,olsen2007evaluating,nebeling2017playing}, and a number of XR toolkits aim to simplify the XR development workflow.
DART~\cite{macintyre2004dart,gandy2014designer} seminally presented foundational concepts and techniques for prototyping AR experiences.
Unity-based toolkits target data visualization (DXR~\cite{sicat2018dxr}, Miria~\cite{buschel2021miria}), remote VR studies (RemoteLab~\cite{lee2022remotelab}), UI adaptation (AUIT~\cite{belo2022auit}), accessibility for blind and low-vision users (SeeingVR~\cite{zhao2019seeingvr}), cross-device interaction (XDTK~\cite{gonzalez2024xdtk}), and distributed collaboration authoring (XSpace~\cite{herskovitz2022xspace}).
DepthLab~\cite{du2020depthlab} provides an ARCore-based toolkit for working with depth across application-specific representations, \revisionchanges{focusing on mobile AR rather than on-headset splatting}, and Palimpsest~\cite{lipkowitz2024palimpsest} anchors spatial UI to physical objects in Vision Pro.\looseness=-1

\splatxr draws inspiration from these works in both their research processes and artifacts: we derive design objectives, outline our foundational concepts and techniques, and then present demonstrative applications. \looseness=-1

\subsection{Gaussian Splatting Tools}
Since the inception of splatting~\cite{kerbl3Dgaussians}, a variety of tools have emerged that seek to satisfy different purposes.

\emph{Training and rendering splats} is at the core of Kerbl et al.'s original implementation~\cite{kerbl3Dgaussians}, Brush~\cite{brush}, and Nerfstudio's gsplat library~\cite{nerfstudio,ye2025gsplat}; these low-level implementations require substantial domain expertise and are incompatible with XR engines such as Unity, Unreal, or WebXR.
Postshot~\cite{postshot}, Luma~\cite{lumalabs},
Kiri Engine~\cite{kiri}, and Polycam~\cite{polycam} simplify training for non-expert users through a higher-level user interface, but are self-contained applications instead of developer dependencies.\looseness=-1

\emph{In-browser viewing and editing of splats} motivates tools such as SuperSplat~\cite{Supersplat,playcanvas}, web renderers~\cite{antimatter15,mkkellogg}, Nerfstudio's viewer \cite{nerfstudio}, Gracia~\cite{graciavr},
Niantic's Scaniverse~\cite{scaniverse}, Spline~\cite{spline}, Sparkjs~\cite{sparkjs}, and Arrival.Space~\cite{arrivalspace}.
Built on the low-level WebGL API~\cite{webgl}, on general-purpose 3D engines such as Three.js~\cite{threejs}, or on special-purpose visualizers such as Viser~\cite{yi2025viser}, these tools support viewing and, in some cases, pre-defined editing; Gracia and SuperSplat go further with stereoscopic viewing, and other web-based experiences reach the headset via the WebXR API~\cite{webxr}.
Their design nevertheless remains centered on viewing, and their browser-based architectures are incompatible with native XR development in engines such as Unity and Unreal.

\emph{Integrating splats into a 3D engine} is facilitated by platform-specific engine plugins: Luma~\cite{lumalabs} and Volinga~\cite{volinga} for Unreal, NVIDIA Omniverse~\cite{omniverse} for robotics simulations, the GSOPs plugin~\cite{gsops} for SideFX Houdini, and ReshotAI's plugin~\cite{gsblender} for Blender.
All of these focus on editing and compositing splats but lack proper tooling for building interactive experiences with them.
\revisionchanges{Pranckevičius' splat renderer for Unity~\cite{unitygs} follows a PC-VR architecture requiring a tethered connection and it offers no XR-specific features, making it unsuitable for standalone deployment. Ninjamode~\cite{kleinbeck_multi-layer_2025ninjamode} likewise provides a Unity splat renderer. Like the authoring tools GaussianShopVR~\cite{shen2025GaussianShopVR} and Splatshop~\cite{schutz12025splatshop}, it offers no splat physics, shadows, or AR blending, and several of these tools remain tethered.}\looseness=-1

In summary, while these tools are useful for different purposes, none addresses the need for building interactive XR applications.
\revisionchanges{To this end, \splatxr introduces a Unity-based toolkit, designed for standalone, untethered headsets including Meta Quest 3 and Apple Vision Pro, that provides real-time performance while leveraging game engine features to enable immersive experiences, simplifies recurring tasks such as preprocessing and importing splats, and provides building blocks for rendering and user-splat interactions.}


\section{Multi-Layered Reality Framework}
\begin{figure*}[t]
\centering
\includegraphics[width=\linewidth]{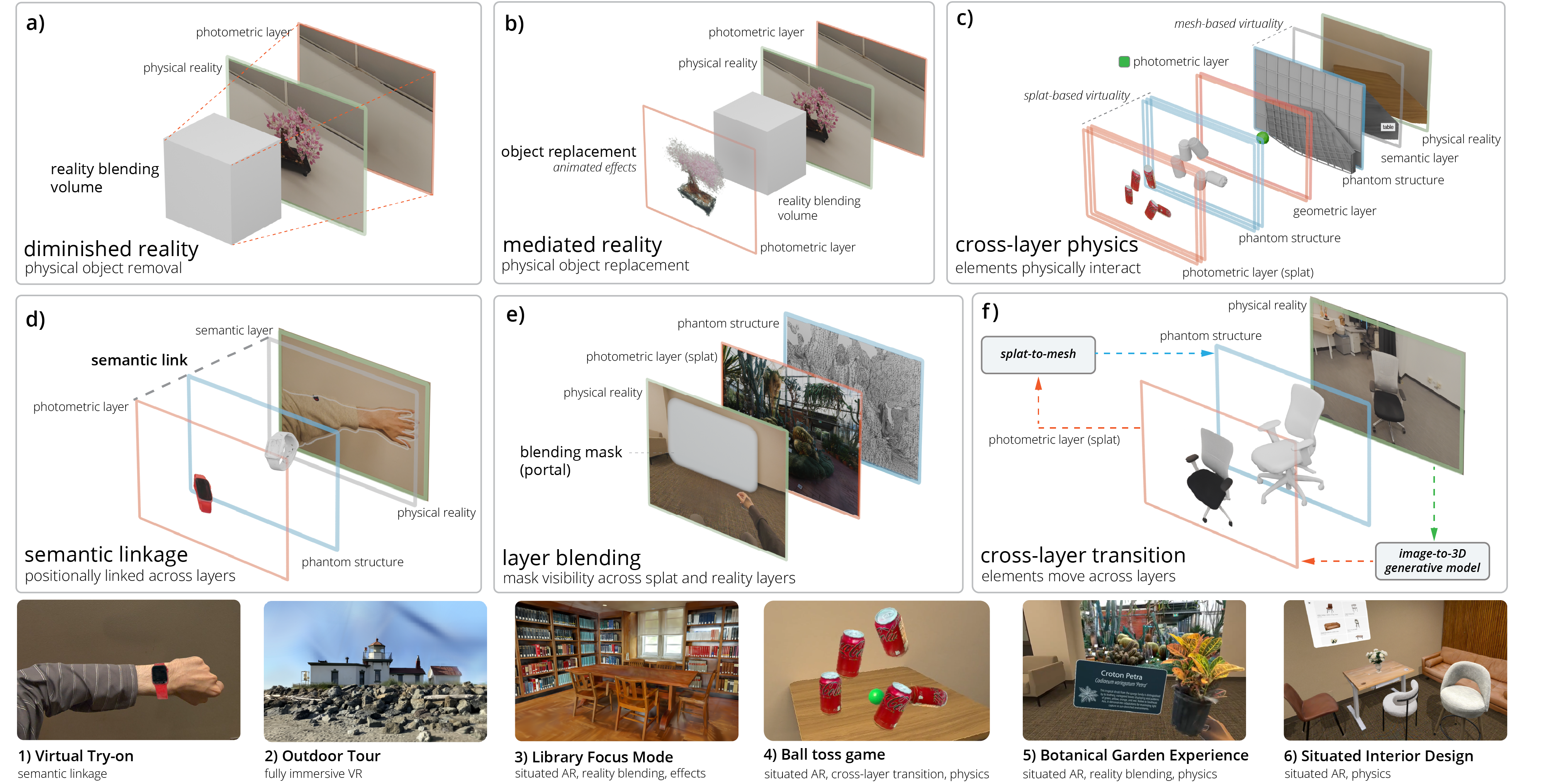}
\Description{Figure 2 has two parts. The top has six subfigures showing isometric layer arrangements. (a) "diminished reality: physical object removal", three layers: a 3D cube ("reality blending volume"), a LEGO bonsai tree ("physical reality"), and the table without the tree ("photometric layer").(b) "mediated reality: physical object replacement", same layers plus a "photometric layer" showing the bonsai mid-disintegration, labeled "object replacement animated effects." (c) "cross-layer physics: elements physically interact", six layers: soda cans mid-air ("photometric layer (splat)"), the cans as grey meshes ("phantom structure"), a green ball ("geometric layer"), meshed wall and table ("phantom structure"), a greyed table annotated "table" ("semantic layer"), and an empty table and wall ("physical reality"). A line between layers 1 and 2 is labeled "splat-based virtuality"; a line between layers 4 and 5, "mesh-based virtuality."(d) "semantic linkage", four layers: an Apple Watch splat ("photometric layer"), a grey watch mesh ("phantom structure"), a hand outline ("semantic layer"), and an image of a hand ("physical reality"). (e) "layer blending: mask visibility across splat and reality layers", three layers: a hand in a room with a grey panel ("blending mask (portal) and physical reality"), a botanical garden ("photometric layer (splat)"), and a garden mesh ("phantom structure"). (f) "cross-layer transition: elements move across layers", three layers: a grey office chair splat ("photometric layer (splat)"), a 3D chair mesh ("phantom structure"), and a photo of the chair in a room ("physical reality"). A boxed label "splat-to-mesh" points from layer 1 to 2; "image-to-3D generative model" points from layer 3 to 2. The bottom shows six application screenshots: (1) "Virtual Try-on / semantic linkage", outstretched hand with a splat watch; (2) "Outdoor Tour / fully immersive VR", a lighthouse on a shore; (3) "Library Focus Mode / situated AR, reality blending, effects", a library and desk splat; (4) "Ball toss game / situated AR, cross-layer transition, physics", soda cans mid-air with a green ball on a table; (5) "Botanical Garden Experience / situated AR, reality blending, physics", a user holding a plant in front of a garden portal; (6) "Situated Interior Design / situated AR, physics", a living room with furniture and a furniture panel.}
\caption{Configurations and operations. \emph{Top:} Different multilayered reality operations across splats, meshes, and reality showcasing cross-layer transitioning, physics, linkage, and blending. \emph{Bottom:} Demonstrative applications that use the above operations to enable XR experiences.}
\label{fig:configs_of_reality}
\end{figure*}

\revisionchanges{Enabling inter-object and human-object interactions between splats, classical meshes, and the physical scene requires systematically managing these entities on both a technical and a conceptual level. 
To this end, we describe a multi-layered framework that conceptualizes reality as a set of logically isolated, yet operationally intertwined layers that align visible and invisible splat and mesh geometry with associated semantics, appearance, and physical context.
In the following, we outline the proposed set of layers and the layer-bridging operations defined on them.}


\subsection{Reality Layers}
Our framework uses five primitive layers:

\subsubsection{Physical layer} The physical layer comprises the user's environment, visually captured and displayed through the passthrough feed of the XR headset.

\subsubsection{Photometric layer} The photometric layer refers to appearance-related components of a scene, such as splats or a rendered frame, which may lack explicit structure. 
Appearance can be obtained from captures of real spaces or generated by a 3D foundation model. 
Its content comes from the rendering engine, which provides the rendered view defined by the headset's intrinsics and extrinsics.

\subsubsection{Geometric layer} The geometric layer includes traditional 3D polygon objects which possess explicit geometry that can be rendered visually while interacting with the scene's physics.

\subsubsection{Phantom structure layer} The phantom structure layer contains invisible explicit geometries, meshes in our case, that are isometrically anchored or aligned to visual layers, letting a visual splat cast a shadow or a physical table support a virtual bouncing ball.
\revisionchanges{Such invisible proxy geometry is a standard graphics technique; we name it as a layer here to make its role in splat-based composition explicit.}

\subsubsection{Semantic layer} \revisionchanges{The semantic layer encompasses a higher-level} understanding of element function and identity in reality via passthrough (e.g., hand tracking or object recognition) to enable cross-layer interactions like linkage and scene understanding.\looseness=-1

\subsection{Operations}
While layers define the static compositional structure of a mixed-reality environment, operations enable the dynamic transitions and interactions between them.
Fig.~\ref{fig:configs_of_reality} depicts possible configurations and operations, which we categorize into three primary types: transformation, blending, and linkage.\looseness=-1

\subsubsection{Cross-layer transformation}
Elements can be transferred across layers (Fig.~\ref{fig:configs_of_reality}f). To move an element from the physical environment to a digital representation, a region of the passthrough feed (physical layer) is processed by a generative model such as SAM 3 or SAM 3D Objects, which lifts 2D visual data into an implicit 3D structure and decodes it into an explicit 3D Gaussian splat (photometric layer).
\revisionchanges{To let particle-based splats interact meaningfully, we rely on a splat-to-mesh operation} which derives an invisible phantom mesh (phantom structure layer) from the splat file and aligns the visual splat to it, yielding a mesh-augmented splat that can physically collide, cast shadows, and interact with classical meshes and other scene layers. \looseness=-1

\subsubsection{Layer blending}
Operations also control how layers are composited within the user's view. Additive operations selectively stack and reveal layers using spatial masks and custom splat blending shaders, as shown in Fig.~\ref{fig:configs_of_reality}e. This enables volumetric portals: the user sees their raw physical surroundings everywhere except within a defined boundary, through which an imported splat environment becomes visible, blending smoothly with the surrounding passthrough as the viewpoint shifts.
Fig.~\ref{fig:configs_of_reality}a shows an example of diminished reality using a reality blending volume covering a real object, and Fig.~\ref{fig:configs_of_reality}b shows an instance of mediated reality, which inserts a modified version of that object into the physical space as a photometric layer.\looseness=-1

\subsubsection{Linkage}
Whereas the operations above are transitions, linkage defines persistent connections that maintain spatial and contextual coherence in a mixed-reality scene by binding digital elements to the physical environment. We categorize these into two types.\looseness=-1

\paragraph{\revisionchanges{Geometric} Linkage} This establishes a spatial relationship between layers, anchoring virtual elements to specific coordinates or distinct physical features. A possible implementation is AR image target tracking, where a virtual splat or phantom mesh is persistently locked to the position and orientation of a physical marker, ensuring static spatial alignment across the physical, geometric, and photometric layers.

\paragraph{Semantic Linkage} Beyond rigid coordinates, semantic linkage dynamically binds a layer to contextual elements in physical reality, requiring a real-time understanding of the user or environment. In Fig.~\ref{fig:configs_of_reality}d, \revisionchanges{hand-tracking data from the physical reality layer links a virtual watch splat to the user's wrist, ensuring continuous positioning and enabling behaviors like dynamic occlusion from the geometric layer.}
\section{\textsc{SplatXR} Toolkit}
\revisionchanges{In the following section, we first outline \splatxr's toolkit architecture and developer experience which together aim to provide developers} with a set of \emph{building blocks} that facilitate developing a splat-enhanced end-user experience. \revisionchanges{We then detail the developer workflow and implementation for each of these building blocks, and finally present \splatxr's capability for facilitating PC-offloaded 3D object reconstruction from the passthrough video feed.}


\revisionchanges{\subsection{Architecture and Developer Experience}
\splatxr is implemented as a Unity package, thus making it easily installable through Unity's Package Manager.
It features a developer-facing graphical interface, integrated into the Unity Editor, that provides access to a set of well-defined building blocks through panels, scripts, materials, and shaders. 
Building blocks can be implemented as needed, and previewed interactively in the Unity Editor scene view. 
Notably, this makes \splatxr's splat renderer compatible with non-XR applications as well.
}


\subsection{Building Blocks}

\noindent \splatxr comprises four main building blocks:
\begin{enumerate}[label=\textbf{\emph{B\arabic*:}}]
    \item a \emph{Splat Renderer} building block enabling splat rendering on XR headsets in passthrough AR or full VR with computational constraints\looseness=-1
    \item a \emph{Splat Physics} building block based on our mesh augmentation procedures, enabling both shadows, cast from splats onto the environment, and rigid-body collisions between splats, classical meshes, and the environment
    \item a \emph{Splat Effects} building block based on customizable shader programs, enabling compelling visual experiences such as object dis- and re-integration
    \item a \emph{Splat Blending} building block enabling blending between the physical scene and the virtual scene in AR
\end{enumerate}


\noindent
Building blocks can be accessed by developers via the \emph{developer interface}, which is \revisionchanges{implemented as a dedicated panel in the Unity Editor. Actions performed in the developer interface add or modify assets in the project folder as well as game objects and attached components in the scene hierarchy. We show our editor interface in more detail in Appendix \ref{app:developer-interface}.} 


\subsubsection{\textbf{\textup{B1: Splat Renderer}}}
The \emph{Splat Renderer} building block comprises a 3D Gaussian splat rendering pipeline designed for compatibility with standalone XR devices, including Meta Quest and Apple Vision Pro. 
On both Quest and Vision Pro, it can render fully immersive splat-based VR scenes, and on Quest, it can render splats and blend them with the passthrough layer for situated AR experiences.\footnote{Apple Vision Pro \revisionchanges{is supported, but passthrough shader functionality is currently limited by Apple's PolySpatial platform.}}
It can be executed untethered on standalone headsets or tethered to a PC. In our multi-layered reality framework, this represents the photometric layer.\looseness=-1

\paragraph{Developer Workflow}
To leverage the renderer, developers simply import a splat into a Unity scene by making use of the \splatxr Importer panel. 
Within the panel, they choose a standard splat \texttt{ply} file, which they either create themselves or download from a third party.
Upon import, the plugin automatically adds a game object to the scene hierarchy, applies a predefined splat material associated with our splat shader, and adds a splat rendering script to the game object.
The renderer not only works properly on-headset but is also aware of when it renders in-editor, switching to the correct editor camera and rendering parameters accordingly.
This ensures faithful in-editor previews of the splat and makes the splat feel like other GameObjects in Unity, allowing for streamlined development.\looseness=-1

\begin{figure}[t]
\centering
\includegraphics[width=\linewidth]{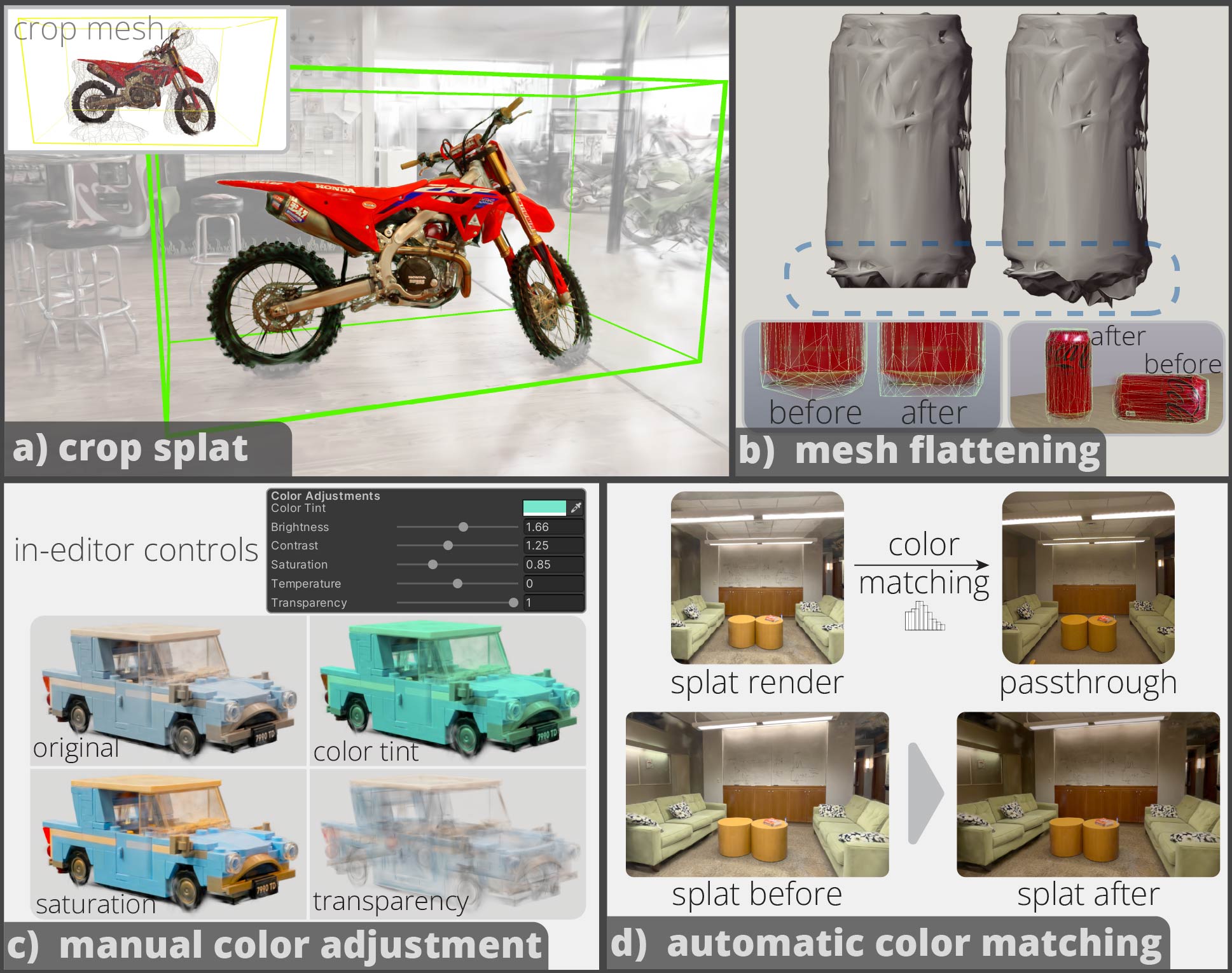}
\Description{Figure 3 shows four sub figures. The first subfigure is labeled “a) crop splat” and shows a motorcycle in a showroom where the showroom background scene is blended white and the motorcycle is bounded by a box with a thick green outline. In the top left corner there is a view of the motorcycle with a mesh superimposed over it with a label as “crop mesh”. The second subfigure is labeled “b) mesh flattening” which shows two soda can meshes side by side. There is a blue border over the bottom parts to highlight that the left can has a flat mesh surface and the right can has an uneven bumpy surface.
There are two smaller subsubfigures. The left one shows the splats of the can bottoms, one showing the collision outline over each one. The right one shows the cans on a table surface, one that is standing upright labeled “after” and one that has fallen down as labeled “before”. The third subfigure is labeled “c) manual color adjustment” which shows an in-editor screenshot with slider controls to edit the splat color properties: “color tint”, “transparency”, “contrast”, “saturation”, “brightness”, and "temperature". Below it there are four views of a LEGO car modified each with coloring transformations of (1) the original car, (2) color tint showing a teal colored car, (3) saturation showing a car with brighter colors, (4) transparency showing a slightly transparent car. The fourth subfigure is labeled “automatic color adjustment” which shows a a before and after shot of a room, one with white color balance labeled “splat” and one with a yellowish tint labeled “target photo”. There is an arrow accompanied by histogram bar icons labeled “color matching”. Below there are two renders of the room showing before and after of the room with similar coloring at slightly different viewpoints.}
\caption{
    Adjustments: a) Cropping of splats and their aligned meshes; b) Planar flattening to fix reconstruction artifacts on flat surfaces; c) Manual control for visual attributes (tint, saturation, and transparency); d) Auto color adjustment to the passthrough environment via histogram matching.}
\label{fig:splat_adj_figure}
\end{figure}

\paragraph{Implementation}
We implemented the renderer as a shader program in Unity's high-level shader language (HLSL). 
The shader program is based on the functionality presented in the original splatting paper~\cite{kerbl3Dgaussians}. 
At design time, once the developer specifies a splat \texttt{ply} file in the import panel, \splatxr converts it into a custom CSV format by performing preprocessing steps that include computing the covariance matrix needed by our renderer, and denormalizing color and transparency information.
At run time, our splat rendering script reads the custom splat file, which contains the attributes of each 3D Gaussian (covariance, color, alpha), and then assembles an arrangement of triangles. 
The triangles are passed into the shader memory and sorted from the perspective of a given camera (either in-editor or on-headset) at an adjustable rate (1 Hz by default) in a background thread on the CPU in the rendering script.
The shader takes care of rendering each splat-associated triangle and alpha-blending it for the rendering result.\looseness=-1


\paragraph{Level-of-Detail Representation}
High-resolution splats lead to noticeable frame drops, and very high-resolution splats, e.g., with 1 million Gaussians or more, become infeasible on current mobile GPUs, such as the Qualcomm Snapdragon XR2 Gen 2 found in the Quest 3, especially because the computational budget needs to be shared with many other tasks that are performed on the device.
To ensure that end-users can still benefit from high-resolution splats, we propose and implement a level-of-detail (LoD) rendering approach.
It displays the high-resolution splat only in close proximity to the user while rendering lower-resolution splats with increasing distance.
In our implementation, we use three levels of detail.
To make use of the level-of-detail representation, developers merely need to provide the three splats with varying splat counts (obtainable by tuning the corresponding parameter in the splat training tool) and select the respective renderer in \splatxr's Renderer panel.

\paragraph{\revisionchanges{Geometric and }Photometric Splat Adjustments}
The renderer also provides developers with additional options to pre-process the input splats' properties by adjusting the Gaussians.
Developers can interactively or programmatically specify a crop box to remove excess geometry surrounding a splat, as shown in Fig.~\ref{fig:splat_adj_figure}a.
Only Gaussians within the crop box are rendered, and the augmenting mesh is jointly cropped.


Overlaying splats on their physical counterpart can lead to a visual mismatch in AR due to the differing lighting conditions of the object at the time of capture and the user's current environment visible through the passthrough. 
As a remedy, \splatxr enables developers to interactively or programmatically adjust a splat's photometric properties in different ways. 
First, they can manually tune color tint, brightness, saturation, contrast, temperature, and transparency via in-editor sliders and color pickers, as shown in Fig.~\ref{fig:splat_adj_figure}c. 
Second, they can provide \splatxr with a passthrough image from the target view and a screen capture of the virtual view.
Using ASC-CDL color grading and histogram matching, our scripts align splat colors for seamless blending, as shown in Fig.~\ref{fig:splat_adj_figure}d.


\begin{figure}[t]
\centering
\includegraphics[width=\linewidth]{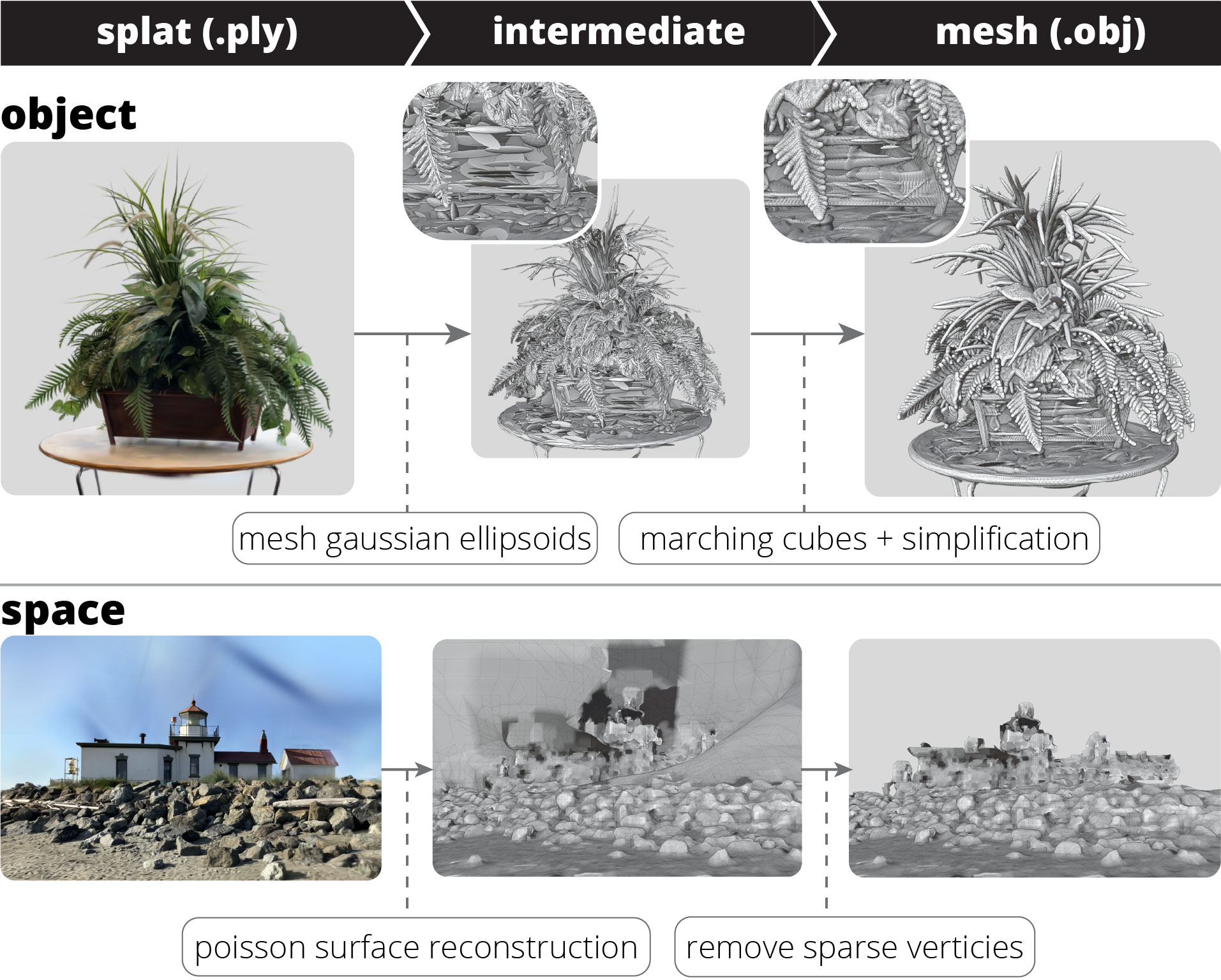}
\Description{Figure 4 is composed of three rows. Each row is composed of three steps. The top row is a header row, showing a chevron process diagram. The steps in the chevrons are labeled first, splat (.ply), second, intermediate, third, mesh (.obj). The second row is captioned "object" and shows the three steps by means of three images. The splat step shows a plant with a lot of intricate detail. The intermediate step shows a grayed out geometry of that same plant, again showing a lot of intricate detail. An arrow connects the first to the second step, and it is labeled "mesh gaussian ellipsoids". A zoom-in annotation on the geometry shows that it is composed of many small ellipsoids. The mesh step shows the same plant, again with a lot of intricate detail. An arrow connects the second to the third step, and it is labeled "marching cubes simplification". A zoom-in annotation on the geometry shows that the plant is now represented by very accurate and fine geometry. The second row is captioned "space" and shows the three steps again by means of three images. The splat step shows a lighthouse at a beach. The intermediate step shows a grayed out geometry of that same lighthouse. An arrow connects the first to the second step, and it is labeled "poisson surface reconstruction". The geometry seems very rough and coarse, and therefore no zoom-in annotation is provided. The mesh step shows the same lighthouse, with the geometry remaining coarse, but cleaner. An arrow connects the second to the third step, and it is labeled "remove sparse vertices".}
\caption{Splat-to-mesh algorithms enable splat physics. \emph{Top:} Marching-cubes-based implementation for objects. \emph{Bottom:} Poisson-reconstruction-based implementation for spaces.}
\label{fig:splat2mesh}
\end{figure}

\subsubsection{\textbf{\textup{B2: Splat Physics}}}
While splats offer astonishing visual renderings, their representation lacks surface information. 
\revisionchanges{This lack of surface information prohibits the direct use of standard graphics models for physics simulations, e.g., for shadow map computations or collision detection.}
We therefore augment the splat with an underlying surface mesh representation that can be accessed by the 3D engine for physics simulations, while we still rely solely on the splat to write to the visual framebuffer, thus obtaining the best of both worlds. In our multi-layered reality framework, this allows us to create a phantom structure from the splat.
We subsume this functionality in the \emph{Splat Physics} block. Fig.~\ref{fig:splat2mesh} provides an overview.

Based on the selected algorithm and parameters, \splatxr creates a mesh in the OBJ format from the splat, automatically co-aligns it with the input splat, parents it to the splat game object, and assigns mesh colliders as needed.
This ensures only the splat is visible, while physics and interactivity take advantage of the surface mesh.
\splatxr implements two mesh generation algorithms.
First, we implemented a custom marching-cubes-based algorithm that produces fine detail but hinges on the assumption of convexity of the splat to be reconstructed.
While this assumption is sufficiently satisfied for objects, it does not hold for spaces, and therefore, we also implemented a Poisson-based reconstruction algorithm for spaces.\looseness=-1

\paragraph{Developer Workflow}
To enable physics simulations for an imported splat object, developers can check the \emph{generate mesh} checkbox in \splatxr's import panel, select an appropriate mesh generation algorithm, and override required parameters as needed. To maintain simplicity in the developer interface, we allow developers to choose between an \emph{object-optimized} reconstruction based on our marching cubes implementation (ideal for close interaction, physics simulations, casting shadows, and manipulation) and a \emph{space-optimized} approach for indoor or outdoor spaces based on the Poisson-based implementation (apt for providing coarser collision meshes for larger environmental scenes).


\begin{figure}[t]
\centering
\includegraphics[width=\linewidth]{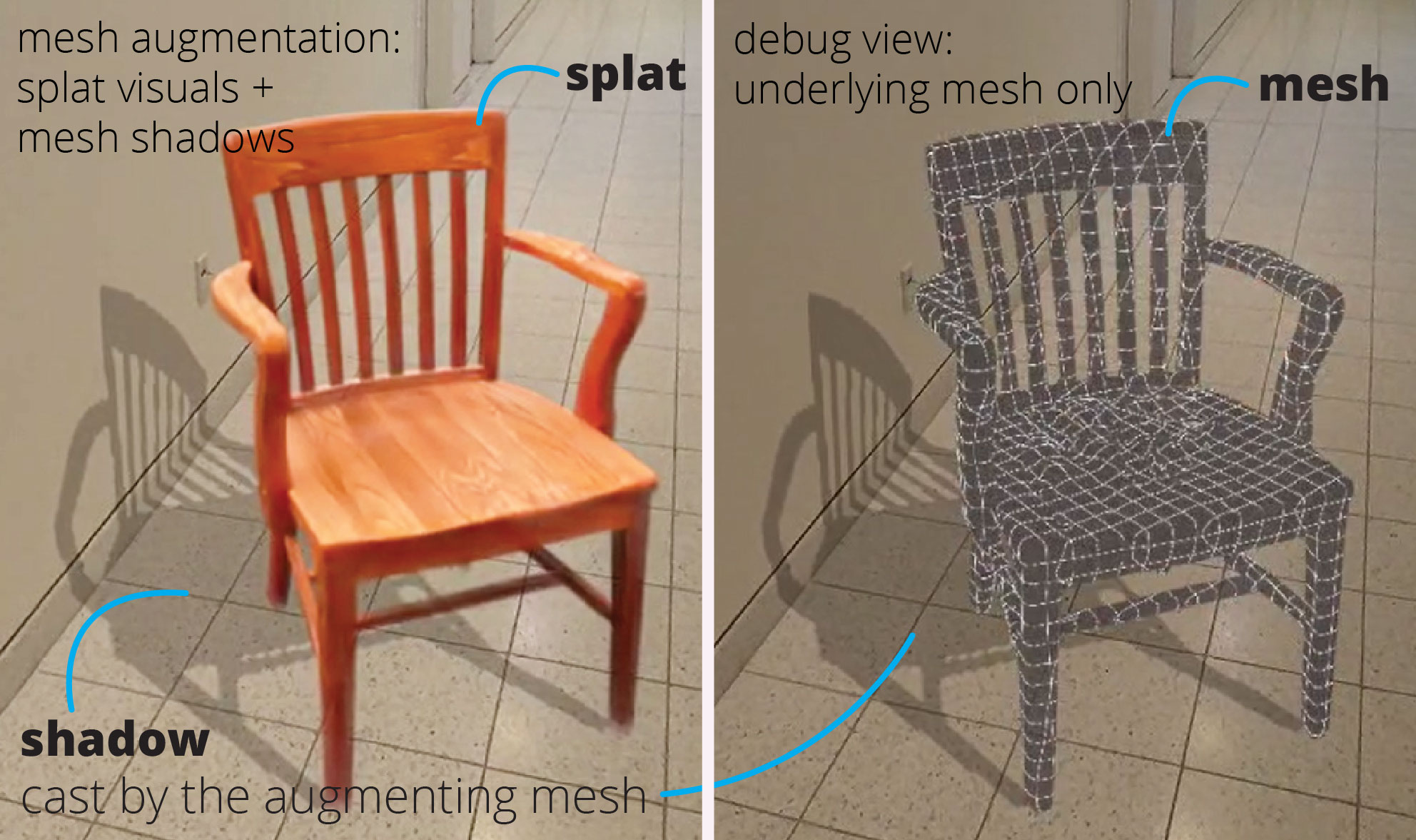}
\Description{Figure 5 is composed of two subfigures. The left subfigure shows an AR screenshot of a physical hallway, augmented by a virtual wooden chair. The chair casts a shadow onto the floor and bends up along the wall as expected. The subfigure is captioned "mesh augmentation: splat visuals + mesh shadows".
A label annotates the chair, reading "splat". A label annotates the shadow, reading "shadow cast by the augmenting mesh". A line extends from the same label to the subfigure on the right. The subfigure on the right shows the same scene, and the chair is still there, but instead of being shown in true colors, the chair is now visualized as a wireframe mesh instead of showing its wooden texture. The subfigure is captioned "debug view: underlying mesh only". The chair bears a label, reading "mesh". The chair again throws a shadow on the floor and wall. The line that was extending from the left subfigure reading "shadow cast by the augmenting mesh" is annotating this shadow.}
\caption{Shadows as part of the \emph{Splat Physics} building block. Shadows can be cast by splat objects onto the physical environment, enabled through mesh augmentation. Collisions, also enabled through mesh augmentation, are supported.}
\label{fig:shadows}
\end{figure}

\paragraph{\revisionchanges{Implementation:} Object-Optimized Phantom Mesh Generation}
To generate a mesh from an object splat, we first create a mesh for each Gaussian ellipsoid from the input PLY file, arranged in a mesh grid. 
The intermediate representation is a mesh reconstruction of each Gaussian, which together represent the entire object. 
Next, a splat grid and an isosurface are constructed based on the densities from the Gaussian ellipsoids. 
We then perform marching cubes over this result to create a mesh with a smoothed surface before decimating it for better performance. 
Object reconstructions can vary in accuracy depending on the structural characteristics of the input splat, and \splatxr allows developers to tune parameters for the splat scale and meshing grid threshold.
Problematically, uneven surface reconstruction at the base of splat objects leads to undesirable behavior in interactions, e.g., toppling over when the object is put onto a flat surface.
Therefore, we also implement a flattening procedure that allows developers to flatten vertices at the base of the mesh along a specified axis. 
Fig.~\ref{fig:splat_adj_figure}b illustrates a can's mesh-augmented splat after flattening, enabling upright placement.

\paragraph{\revisionchanges{Implementation:} Space-Optimized Phantom Mesh Generation}
For large-scale reconstructions, we employ the Poisson surface reconstruction algorithm, and then crop the results by removing vertex outliers to efficiently produce large-scale meshes. 
We use a coarser resolution for scenes than for individual objects to manage compute at scale. 
This approach recovers the largest surfaces, e.g., the floor, walls, and enclosing sides, needed for collisions.



\paragraph{Shadows and Occlusions}
Splat objects that have been augmented with the generated and co-aligned mesh are registered with Unity's native shadow mapping procedure, enabling them to cast shadows onto other objects in the scene.
On Quest devices, \splatxr leverages the Meta Scene API to obtain an environment mesh, automatically configures it as a phantom shadow catcher \cite{breen1996interactive,kalkofen2011visualization}, and thus enables the visual impression of casting splat shadows onto the physical world.
Practically, \splatxr automatically attaches a custom shadow-casting material to the augmenting mesh.
The shadow-casting material uses an HLSL program that writes to neither the framebuffer nor the depth buffer---leaving the augmenting mesh invisible and non-interfering with the splat renderer---while still contributing a shadow-casting pass to Unity's shadow mapping. \revisionchanges{Fig.~\ref{fig:shadows} shows a shadow of a splat on a physical environment.
Occlusions where splats cover parts of an otherwise visible mesh are implemented by culling occluded mesh geometry in the splat shader.}\looseness=-1

\paragraph{Collisions}
\splatxr further makes use of the augmenting mesh for collisions.
To this end, it automatically attaches a mesh collider to the augmenting mesh, thus computing a tightly fitting convex hull over the vertices within that mesh.
This in turn allows attaching Unity's rigid-body behavior to the object for collision simulation.
Fig.~\ref{fig:configs_of_reality} shows an example of a splat object colliding with other splats (coke cans), other classical mesh objects (green ball), and the physical environment (table).
Based on the rigid-body behavior, \splatxr can automatically enable grab interactions with the splat.
Furthermore, when used with Unity's navigation mesh component on a space splat, this approach can turn the space splat into a walkable surface for end-users or virtual characters.

\subsubsection{\textbf{\textup{B3: Splat Effects}}}
\begin{figure}[t]
\centering
\includegraphics[width=\linewidth]{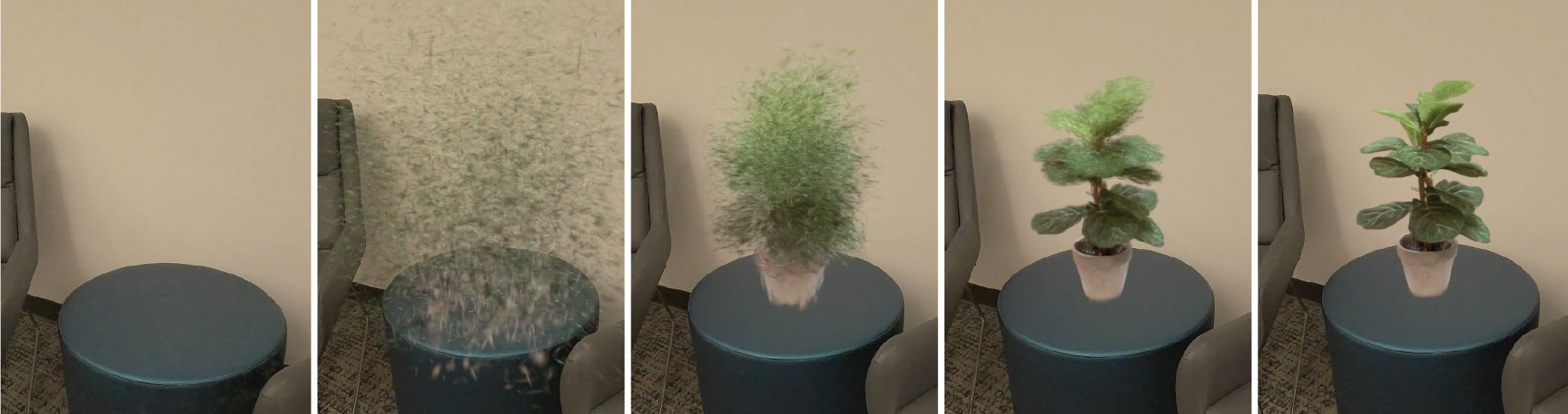}
\Description{Figure 6 shows five sub panels each displaying a keyframe of an animated visual re-integration effect of a potted plant splat sitting on a table. The first panel shows an empty table, and the following panels show a cloud volume forming into points, and finally into the plant.}
\caption{
   Object reintegration effect as part of the \emph{Splat Effects} building block.
}
\label{fig:object-reintegration}
\end{figure}

The smooth alpha falloff and the compositional characteristics of a splat lead to an ``organic emergence'' of its geometric structure as part of the optimization process. 
In contrast, the structure of a mesh is constructed rather mechanistically from the underlying point cloud.
In the \emph{Splat Effects} building block, we leverage this property of splats to our advantage for vivid animations, as shown in Fig.~\ref{fig:object-reintegration}.

\paragraph{Developer Workflow}
To build a custom splat effect, developers can create a new HLSL shader program by instantiating \splatxr's \texttt{Splat\-Effect\-Shader} template.
This provides the developer with an empty template within which all elementary splat rendering methods, including \texttt{Compute\-Splat\-World\-Centers} and \texttt{Compute\-Splat\-World\-Corners}, are callable as library functions.
Developers can insert custom shader code or override existing shader functions for maximum flexibility, and can access a global animation time index.

\paragraph{Implementation: Splat Dis- and Reintegration}
We demonstrate the effectiveness of this library-templating approach by replicating Andrioli's splat dis- and re-integration effect \cite{andrioli2025splatshadereffects} shown in Fig.~\ref{fig:object-reintegration} in 12 lines of code (LoC), the same 8 LoC as the original implementation, plus 4 additional lines to improve XR integration by fading individual Gaussians into passthrough with increasing distance from their true pose. We created a collection of four additional splat animation effects: radial reveal, pulse, digital rain, and warp.

\subsubsection{\textbf{\textup{B4: Splat Blending}}}
A recurring challenge in XR applications is blending a virtual scene with the physical scene, e.g., to provide real-world awareness \cite{hartmann2019realitycheck}, diminish or animate physical objects from the user's perceived experience \cite{kari2023scene}, or control the degree of immersion \cite{cools2022blending,valkov2017smooth}.
\splatxr facilitates building such applications by providing the Unity elements necessary to render the scene onto any desirable geometry. \looseness=-1

\paragraph{Developer Workflow}
Developers can simply drag the \texttt{Virtual\-Eye\-Material} onto any ``splat-revealing'' mesh geometry in the scene, and tag the splats to be masked accordingly.

\paragraph{Implementation: 3D View-Dependent Portal}
We demonstrate the ease of use of this approach by implementing the 3D portal shown in Fig.~\ref{fig:teaser}, where the virtual Gaussian space splat is rendered onto the physical wall.
The splat is invisible \emph{unless} viewed through the splat-revealing portal.
Because the material can be applied to any custom mesh geometry, developers retain full control over where the mesh appears, its size and shape, and how it changes over time (e.g., for animations).
\splatxr provides the virtual eye material as part of its material library. 
The material is shaded by a \texttt{Virtual\-Eye\-Renderer} that samples the stereoscopic render texture, provided by two eye-aligned cameras, in image space.
By allowing the developer to customize the rendering layers, \splatxr can ensure that only the desired virtual splat geometry is rendered.




\subsection{\revisionchanges{Headset-Guided} Object Reconstruction}
\revisionchanges{\splatxr's building blocks provide a fully self-contained library of reusable components for on-device splatting. 
To bring the advances of compute-intensive vision-foundation models for on-the-fly splat reconstruction to interactive XR applications on compute-constrained headsets, we also provide an edge-computing-based} system for lifting physical objects into splat replicas using image-to-3D foundation models.

To reconstruct an object, users select a single point or multiple points in their physical space through their headset. The system captures the current passthrough view and projects the 3D points into 2D coordinates to serve as prompts for SAM 3 \cite{carion2025sam3segmentconcepts}. The resulting image mask and passthrough frame are processed by SAM 3D Objects \cite{sam3dteam2025sam3d3dfyimages}, which generates a 3D splat within minutes. \revisionchanges{Despite this latency, we expect it to be an acceptable tradeoff for developers as a means of quick 3D reconstruction.} \looseness=-1

To ensure correspondence with the physical world, the object is automatically scaled to life-size using the Meta Quest Depth API by measuring the real-world distance from the headset to the point of selection and appropriately scaling the generated splat from the estimated world coordinates. \revisionchanges{The headset sends image and point data wirelessly to a custom remote server and receives a generated splat file.} The current implementation runs all generative models locally on an edge computer.

\begin{figure}[t]
\centering
\includegraphics[width=\linewidth]{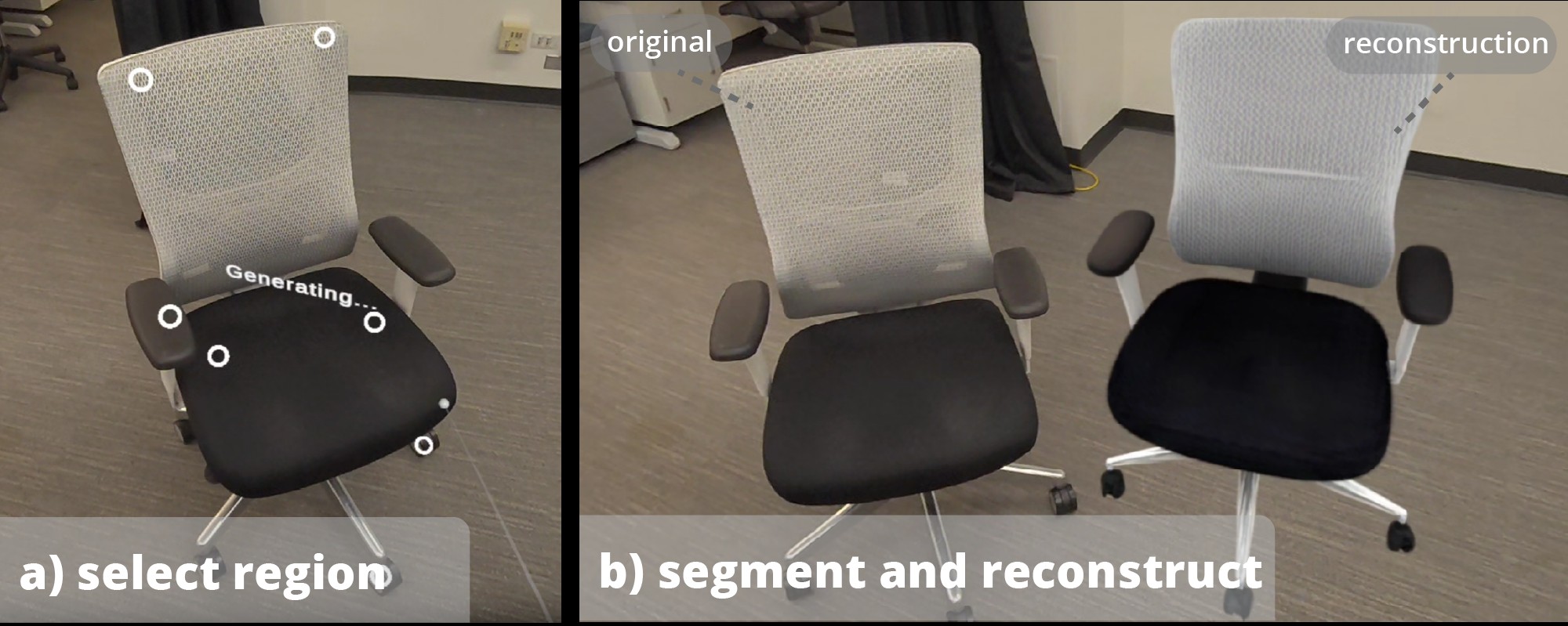}
\Description{Figure 7 has two subpanels. The left panel is labeled "select region" and shows a grey office chair with small points on the chair. There is a text box that says "Generating..." in the middle.
The right panel is labeled "segment and reconstruct" and shows the same chair, labeled "original", and on the right to it is a splat reconstruction of the object with looks similar but slightly lighter colors, labeled "reconstruction".}
\caption{
   Offloaded object reconstruction via headset capture. \emph{Left}: 3D points selected on the object of interest. \emph{Right}: The object is reconstructed and placed next to the original object.
}
\label{fig:object-reconstruction}
\end{figure}


\section{Applications}
To provide an overview of \splatxr's use cases, we developed several applications demonstrating multi-layered reality configurations \revisionchanges{built from \splatxr's building blocks, as shown in Fig.~\ref{fig:configs_of_reality}.
Three capabilities that no prior tool provides together recur across all of them: untethered on-headset splat rendering, which lets them run outside without a rendering server or PC-VR cable; automatic splat-to-mesh proxy generation, which supplies the collision and shadow geometry that splats lack; and AR passthrough compositing with occlusion from environment geometry.}\looseness=-1

\paragraph{Virtual Try-On} An application showcasing \emph{semantic linkage} where a splat of a watch (\emph{photometric layer}) is dynamically overlaid onto the user's wrist (\emph{physical layer}) as identified by the \emph{semantic layer}. \revisionchanges{By exposing the splat as a Unity GameObject, \splatxr's \emph{Splat Renderer} allows developers to parent it directly to the Meta SDK's hand-tracking anchor. This demonstrates how splats can be driven by physical reality and leverage mesh augmentation in \emph{Splat Physics} for occlusion, sparing developers from implementing complex rendering and mesh-occlusion logic.}

\paragraph{Outdoor Tour} A full overlay application that places the user entirely within a captured splat environment, replacing the \emph{physical layer} with a comprehensive splat \emph{photometric layer} \revisionchanges{enabled by the \emph{Splat Renderer}. The user's physical movement is restricted by a \emph{phantom structure} mapped to the splat using \emph{Splat Physics}. Its splat-to-mesh proxy generation yields a mesh collider that prevents users from walking or clipping through surfaces, keeping navigation bounded within physical limits.}\looseness=-1

\paragraph{Library Focus Mode} A \emph{reality blending} application where the user replaces their busy peripheral surroundings with a quiet, static splat of an empty library (\emph{photometric layer}) \revisionchanges{rendered by the \emph{Splat Renderer}}, keeping only their immediate desk visible through passthrough---an instance of diminished and mediated reality that retains grounding in the physical space. \revisionchanges{\splatxr's \emph{Splat Blending} block provides a spatial masking volume around the desk along with shaders that render passthrough inside the volume and the library splat outside it.}

\paragraph{Ball Toss Game} An AR game in which a user tosses a virtual ball at an arrangement of splat soda cans resting on a physical table. Mapping each component to its respective layer (Fig.~\ref{fig:configs_of_reality}c) defines how they interact across different representations of reality: the mixed-reality passthrough forms the underlying base in the \emph{physical layer}, the splat cans are situated in the \emph{photometric layer} \revisionchanges{using the \emph{Splat Renderer}}, and the virtual ball operates within the \emph{geometric layer}, which encapsulates both its explicit geometry and its visual rendering. Cross-layer dynamics, such as the ball bouncing off the table and colliding with the cans, rely on \emph{phantom structures} \revisionchanges{operationalized by \emph{Splat Physics}}. These act as invisible, anchored geometries related to both the physical environment mesh \revisionchanges{(via the Meta SDK scene mesh)} and the splat cans, which underpin the visible elements and allow physics interactions across layers. This decomposition provides a mental model for how distinct visual, structural, and physical elements converge in a unified mixed-reality environment.

\paragraph{Botanical Garden Experience} A volumetric portal, created with additive layering and spatial masking \revisionchanges{from \emph{Splat Blending}}, reveals a botanical garden (\emph{photometric layer} \revisionchanges{via the \emph{Splat Renderer})}. Users can reach through the portal to extract individual plant splats and place them into their physical room (\emph{physical layer}). \emph{Phantom structures} via \emph{Splat Physics}, attached to the garden, plants, and physical environment, allow the splat plants to cast accurate shadows over real-world surfaces and enable collisions across layers. \looseness=-1

\paragraph{Situated Interior Design} Users place high-fidelity furniture splats (\emph{photometric layer} \revisionchanges{via the \emph{Splat Renderer}}) into their physical room (\emph{physical layer}). To ensure plausible interaction, \revisionchanges{\emph{Splat Physics}} generates \emph{phantom structures} for both the imported furniture and surrounding surfaces. This enables physical collisions and realistic shadow casting. \revisionchanges{\splatxr handles proxy generation, environment occlusion, and passthrough color correction so captured furniture matches the room's lighting—an AR-specific challenge unaddressed by VR-only systems.}

\revisionchanges{\section{Technical Evaluation}
We evaluated our renderer and toolkit to investigate different components of the system and their support for a range of splat types for standalone experiences. 
We tested 27 configurations: 3 objects (plant, car, bonsai), 3 indoor scenes (office, botanical, library), and 3 outdoor scenes (sundial, lighthouse, plaza), each at 3 splat counts (50k, 100k, and 200k). We include additional details and data in Appendix \ref{app:technical-eval} and supplementary materials.

\subsection{Renderer}
Because visible splat density affects performance, indoor and outdoor scenes were measured at both their densest and sparsest viewpoints. Frame rate was capped at 72 fps. 
CPU frame time was negligible across all conditions (M=2.3 ms, SD=0.3 ms), making the results GPU-bound. Metrics were averaged over 3 repetitions on a standalone Quest~3, measured with the OVR Metrics Tool, under Unity default quality, on a full battery, from a fixed position.

For objects, the renderer sustained a full 72 fps for both
50k and 100k splats (GPU time 7.6 and 9.7 ms, with negligible frame drops). At 200k, GPU time averaged 15.4 ms and frame rate averaged 59 fps, with frame drops averaging under 6\%. For outdoor scenes at the sparsest viewpoint, the renderer held 72 fps through 100k and averaged 64 fps at 200k (GPU time 3--11 ms); at the densest viewpoint it averaged 72 fps at 50k, 53 fps at 100k, and 38 fps at 200k (GPU time 12--26 ms), with frame drops staying low. For indoor scenes at the sparsest viewpoint, the renderer averaged 63 fps at 50k, 61 fps at 100k, and 58 fps at 200k; at the densest viewpoint it averaged 58 fps at 50k, 46 fps at 100k, and 36 fps at 200k with GPU time ranging 13--32 ms, with GPU utilization at 92\% and frame drops averaging 2\%.
On launch, memory was 1.3 GB at the first splat, increasing by 60--140 MB per additional 50k splats.
Repeating the full suite reproduced nearly identical results. We observed no performance drop when splat objects were moved or augmented with meshes.
Internal tests mixing objects, scenes, and passthrough showed that performance depends on the total number of visible splats. We also tested our system with combined splats at different scales and with XR scene interactions.

The average scene import and CSV conversion times for the 9 splats increase with the number of splats, measuring 0.73 $\pm$ 0.14 s for 50k, 1.38 $\pm$ 0.12 s for 100k, and 2.5 $\pm$ 0.23 s for 200k splats.

\subsection{Splat-to-Mesh}
We ran splat-to-mesh conversion on the same 27 splat files used in the rendering evaluation. 
The object algorithm was applied to the 3 objects and the space algorithm to the 6 scenes.
Object meshing averaged 25.7 s for 50k splats, 29.9 s for 100k, and 41.4 s for 200k, scaling gradually with splat count.
Space meshing was faster, averaging 18.8 s for 50k splats, 22.7 s for 100k, and 31.6 s for 200k.

\subsection{Object Reconstruction}
This pipeline, which offloads SAM 3D to a PC, reconstructed an object in an average of 2.38 min (SD = 0.16 min) across four trials.

\subsection{Failure Cases}
With our renderer, performance degraded with increased camera proximity and the associated rise in visible splat density. Large high-density splats close to the camera caused stuttering. 
For our splat-to-mesh operation, the coarse space meshing may struggle with scenes containing many detailed objects or intricate structures, as mesh simplification loses this detail.    

}

\section{Preliminary Developer Study}
\revisionchanges{We evaluate \splatxr following Ledo et al.'s~\cite{ledo2018evaluation} framework, pairing a \emph{demonstration} of applications with a \emph{usage} evaluation in the form of a first-use study structured around Olsen's UI systems criteria~\cite{olsen2007evaluating}: importance and STU context (Situations, Tasks, and Users), expressive leverage and match, power in combination, generality, and scale.}
We conducted four expert interviews during the development process of the building blocks and a small \revisionchanges{first-use} study (n=3) after creating an initial version of the toolkit.

\subsection{Formative Expert Interviews}
We conducted four semi-structured interviews with experts from the XR domain and the computer vision/graphics domain to understand their potential needs.
The first participant (E1) was a filmmaker of Unity-based VR films who uses Nerfstudio, Postshot, Blender, and Houdini to train, edit, and render splats as part of their artistic workflow.
The second participant (E2) was a computer vision researcher working on splatting algorithms. The third participant (E3) was an HCI researcher who works with splats and XR. The fourth participant (E4) was a computer vision researcher with experience in neural rendering. 
We first asked a set of basic questions regarding their use of splatting, and then aimed to elicit thoughts on pain points for the problem of bringing splats into XR. \looseness=-1

E1 stressed integration challenges, noting that ``the biggest pain point is that splats are just not playing nicely with everything else in the Unity project'' and wishing that features spread across many tools could be combined into ``one mega tool.'' 
E2 emphasized the need for a real-time preview in any splatting system to ``get a sense of the [splat] density.''
E2 further highlighted the computational efficiency of splats over neural radiance fields, but felt unsure about the feasibility of implementing splat rendering on a mobile GPU.
E1 pointed out that splats really ``shine'' compared to other methods in terms of photorealism of lighting; however, this advantage would be invalidated if even basic lighting phenomena such as shadows were not properly rendered in XR. E1 and E3 mentioned the issue of blending and compositing multiple splats in a scene with current tools. E4 described the need for a ``lightweight'' real-time renderer. E3 described a limitation of splats compared to meshes: ``[if] I want some physics... just throwing this object... mesh will be better.'' E3 found current tools difficult to use and expressed interest in a toolkit that can ``accelerate that process or make it simpler for people to use.'' E4 mentioned that, despite splats' lack of accurate ``3D structure,'' meshing would enable ``physics interaction... that is probably good enough.''\looseness=-1 

\revisionchanges{Using Olsen's strategy~\cite{olsen2007evaluating}, these interviews establish the \emph{importance} of the problem and its STU context: the users are XR developers, the task is assembling XR experiences that combine captured splats with virtual meshes, and the situation is on-headset development for standalone XR where compute is constrained and workflows must integrate with existing Unity/Meta SDK pipelines. The interviews also surface a \emph{problem not previously solved}:} splats are incompatible with Unity projects and difficult to build XR applications around. \revisionchanges{From this, we derive two criteria the toolkit must satisfy: a match with existing XR development practices, and splat-mesh augmentation to enable physics and shadows.}

\subsection{Small-Scale Developer Study}
To evaluate the usability and utility of \splatxr's building blocks, we conducted an IRB-approved remote user study with Unity developers (n=3) who had experience developing XR applications with the Meta Quest SDK. After a 30-minute onboarding, each participant built an XR application of their choice over up to a week, followed by a semi-structured interview, \revisionchanges{which we interpreted through Olsen's heuristics~\cite{olsen2007evaluating}.}

\begin{figure}[H]
\centering
\includegraphics[width=\linewidth]{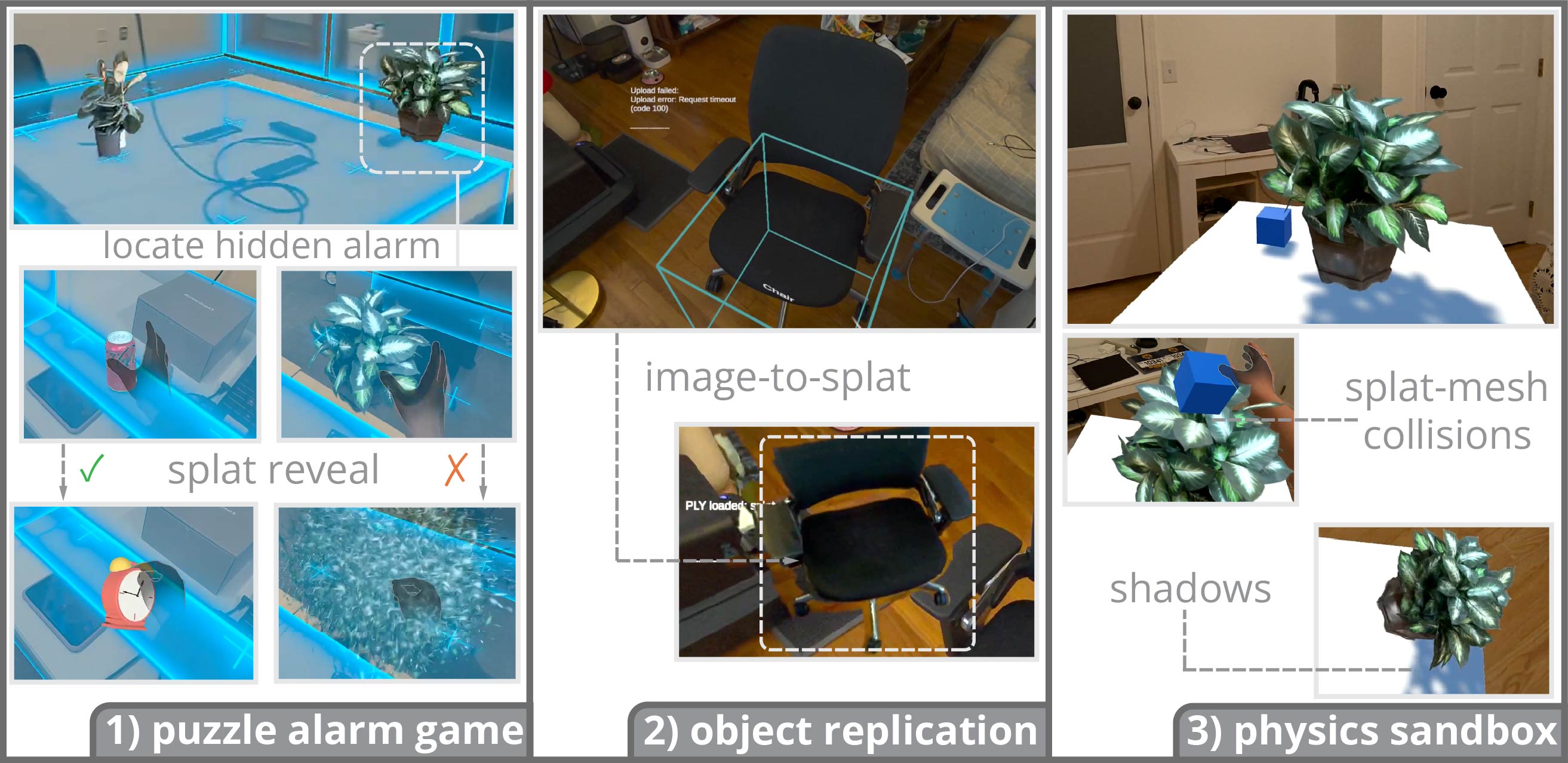}
\Description{Figure 8 shows 3 subpanels. The first subpanel is labeled “1) puzzle alarm game” and shows two small plants on a table where the room has all blue holographic surfaces. Underneath it there is a label “locate hidden alarm”, below that is two images one of a user about to grasp a can and one grasping a plant. Below the can image, there is an arrow with a green checkmark  and below the plant image there is a red X, and in between there is a label “splat reveal.” The can image arrow points to an alarm clock object, and the plant image points to a dissipated plant. The second subpanel labeled “2) object replication” shows a user selecting a chair by pointing with a VR laser pointer showing a bounding box over the chair. There is an arrow out of that pointing to a second sub image of a splat reconstructed chair with a text label “image-to-splat”. The third subpanel is labeled “3) physics sandbox” and shows an image of a large potted next to a blue cube, both on a white surface. There is a smaller image below it with an arrow pointing labeled “splat-mesh collisions”, and another image labeled “shadows” with an arrow pointing to the shadow on the surface.}
\caption{Experiences created by participants using \splatxr.}
\label{fig:dev_apps}
\end{figure}

\revisionchanges{\emph{Generality and Demonstrations.}} Three developers explored diverse applications (Fig.~\ref{fig:dev_apps}): P1 an audio hide-and-seek game whose splat targets disintegrate and are swapped for an alarm clock; P2 a remote task-guidance system reconstructing objects from passthrough photos and rendering them at runtime with pulsing effects; and P3 a physics sandbox scene combining an imported splat with mesh physics, a looping plant effect, and a blending portal. \revisionchanges{The applications built by participants, in combination with our sample applications discussed earlier, support a \emph{generality} claim beyond a single application or use case.}

\revisionchanges{\emph{Expressive Leverage and Match.} All three cleared the toolkit's \emph{threshold} quickly.} P2 installed it in less than a minute (``plug and play''), P3 integrated it in under ten minutes, and P1 viewed an in-headset splat render in about fifteen minutes. \revisionchanges{The \splatxr building blocks also showed \emph{expressive match}}: P3 found them ``really easy and intuitive'' because they mirror the Meta building blocks they already use, and P1 noted a developer's ``intuition... can transfer.'' The toolkit expresses splat capabilities in the vocabulary of Meta's SDK, which XR developers already use. Participants could easily incorporate desired splat capabilities such as meshing or effects, \revisionchanges{reflecting the toolkit's \emph{expressive leverage}}.

\revisionchanges{\emph{Power in Combination.} Developers valued composing splats with meshes, effects, and passthrough using the blocks: the automatic splat-to-mesh step let P2 add shadows and effects that their prior renderer could not, and let P3 make regular splats interactable. Participants rated the renderer and physics/effects blocks most useful;} P2 said the blocks make the toolkit ``easier to pick up completely from scratch.''\looseness=-1

\revisionchanges{\emph{Problem Not Previously Solved and Importance.}} P2 contrasted \splatxr with a prior Unity splat renderer~\cite{unitygs} that was ``incompatible with a lot of Unity versions,'' whereas \splatxr worked ``straight out of the box.'' All participants saw value in \splatxr for their future research: P1 in simulating real-world contexts hard to stage in reality, P2 in photorealistic remote task guidance where ``the photorealism of the splat... matters a lot more than just some approximate mesh,'' and P3 in interacting with whole environments.

\emph{Limitations.} Participants noted where the toolkit's functionality fell short: P3's scene ``glitch[ed] and freeze[d]'' after a portal added a second splat room, and P1 hit unstable physics in large scenes. P2 had to add runtime splat loading, which the package lacked. P1 wanted shader customization and relighting; both P1 and P3 asked for clearer effect configuration and easier base-flattening, \revisionchanges{which motivates future work on rendering and runtime asset handling.}

\section{Limitations and Future Work}
While \splatxr was designed to help developers solve recurring challenges in AR/VR, future research is needed to improve it.

\paragraph{Improving the Renderer}
With our LoD representation, we can render scenes across an order of magnitude (e.g., 100K--1M Gaussians).
However, to enable splats with even \revisionchanges{higher counts}, research into potential computational improvements of the renderer is needed, e.g., by adapting techniques for foveated rendering. Furthermore, \splatxr supports static splats, but a sequence of static splats is less expressive and computationally less efficient than a dedicated representation for dynamic splats. Therefore, it would be worth investigating options to bring dynamic splats to the headset, such as 4D video or splat-based avatars. Additionally, while our framework supports shadows, the splats themselves remain unresponsive to environmental lighting. Exploring real-time relighting techniques on splats would enable more realistic compositing.




\paragraph{Analyzing Splats for Interactive Purposes}
\splatxr abstracts away from the content of a splat.
It would be interesting to add methods based on either standard geometric processing or machine learning to semantically analyze an imported splat and set it up for common uses in XR.
For example, the walkable space of a splat could be derived to navigate a game character on it, to facilitate VR illusions, or to automatically align the splat with the physical walkable area.\looseness=-1

\paragraph{Providing Training Capabilities}
We implemented \splatxr to support developers in integrating existing splats in their XR applications. We believe that it could be worth exploring interactions and system components needed to capture a splat from the XR device's cameras, potentially using generative models.

\paragraph{Outlook}
We believe that a fundamental mismatch has emerged: while established XR pipelines are explicitly geometric, emerging scene representations are increasingly implicit and lack structure.
Because 3DGS lacks the explicit properties of traditional polygonal meshes, it remains isolated from standard graphics techniques like shadows and collisions. Furthermore, unlike 2D pixel-based generation or generative video, XR is defined by 6DOF movement through 3D physical space. This spatial nature demands explicit grounding; we cannot yet operate solely in pixel space when user perspectives and hand interactions are tied to physical 3D coordinates. While recent action-conditioned generative models allow world exploration via keyboard inputs (WASD) \cite{genie3}, such paradigms are fundamentally incompatible with the fluid spatial input of XR.

We designed our framework to bridge this divide by treating 3DGS strictly as a visual rendering layer, an interface that maps camera poses to pixel outputs. In this capacity, 3DGS serves as a performant proxy for the long-term vision of personalized generative \emph{situated world models}, world models situated in reality which will eventually model immersive 3D scenes and dynamics based on an implicit understanding of the physical environment and the user. By decoupling the visual output from the underlying scene structure, our architecture remains representation-agnostic. Whether the underlying representation is 3DGS today or structured fully implicit world models in the future, our system provides a vision for an architecture to integrate these unstructured, implicit 3D representations with the explicit structure required for physical interaction.\looseness=-1



\section{Conclusion}
We presented \splatxr, a Unity \revisionchanges{toolkit that brings Gaussian splatting to standalone XR headsets through modular building blocks: a real-time on-device renderer, splat-to-mesh reconstruction, splat physics, splat blending, and offloaded object reconstruction from headset capture. Together, these components integrate computer vision and graphics techniques into a single developer workflow, making splat-based XR development accessible to the wider XR community.}
By organizing its components around a multi-layered composition of physical and digital primitives, \splatxr allows pixels and objects to move across the boundaries between physical and virtual representations, so that virtual splats cast shadows on real surfaces, physical geometry constrains splat physics, and captured objects re-enter the scene as interactive digital twins. \revisionchanges{Unlike prior splatting work confined to VR, \splatxr equally supports AR, running wire- and server-free with passthrough compositing, alpha blending, and occlusion from reconstructed geometry.}
By exposing these capabilities through a familiar Unity interface, \splatxr provides the technical foundation and mental model to treat the physical and digital as a cohesive, multi-layered reality, where the transition between physicality and virtuality is architecturally supported rather than technically disjointed. \revisionchanges{We release \splatxr as open source to support continued community development.}\looseness=-1
\revisionchanges{\begin{acks}
We thank Szymon Rusinkiewicz for advice and assistance with the meshing algorithm, Simon Bethke for providing the botanical garden dataset~\cite{bethke_botanical_america}, and members of the $\psi$ Lab for their advice. This research was supported by the Princeton NextG Innovation Grant.
\end{acks}}

\bibliographystyle{ACM-Reference-Format}
\bibliography{References,references_mo}
\revisionchanges{\clearpage
\appendix

\twocolumn[{
\section{Appendix}
\vspace{-0.5em}

\begin{minipage}{\textwidth}
    \centering
    \includegraphics[width=0.88\linewidth]{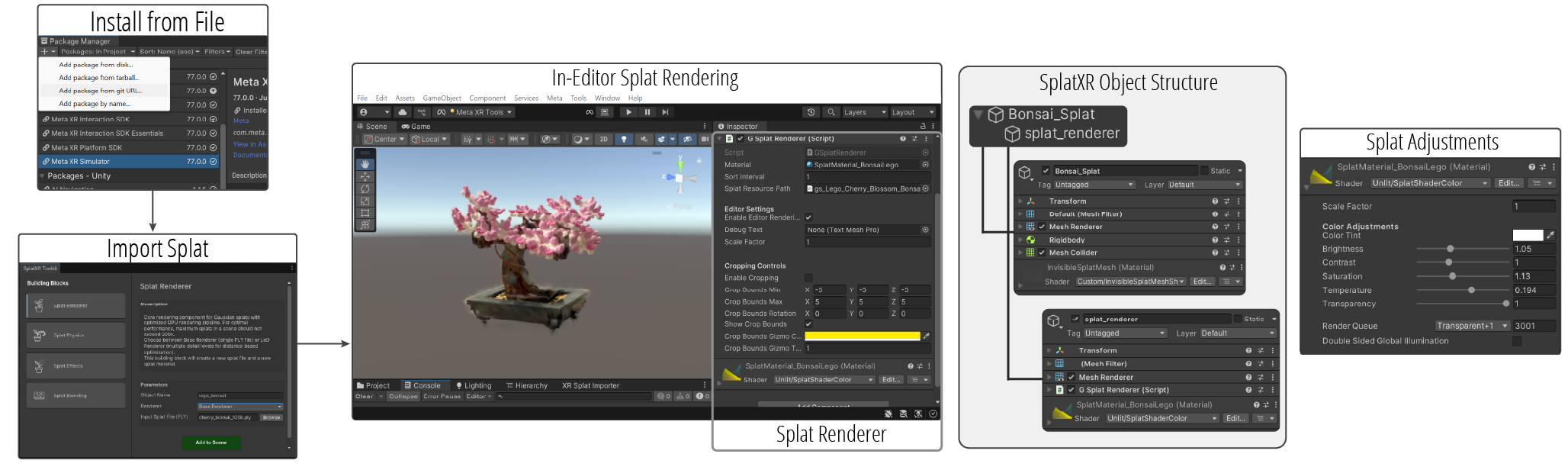}
    \Description{Figure 9 showcases the developer interface screenshots from the Unity editor. On the top left, there is a box labeled “Install from File” which shows the Unity Package Manager and the selection to import the package. An arrow points to a box labeled “import splat” which shows the SplatXR toolkit panel showing a field to input the ply file, a side panel of building blocks, and an Add to Scene button. An arrow points to a box labeled “In-Editor Splat Rendering” showing the Unity editor with a splat in the view with a box around the Inspector showing the renderer component. There is a box on the right labeled “SplatXR Object Structure” showing the hierarchy and GameObjects for the splat. On the right, there is a box labeled “Splat Adjustments” showing the splat material and shader controls.}
    \captionof{figure}{\splatxr's developer interface.}
    \label{fig:devExperienceFigure}
\end{minipage}
\vspace{0.5em}
}]

\setlength{\intextsep}{7pt plus 1pt minus 1pt}

\subsection{Developer Interface}
\label{app:developer-interface}
Fig.~\ref{fig:devExperienceFigure} shows an overview of the developer workflow in \splatxr.
\paragraph{Install from File}
Developers can import the toolkit package downloaded from our repo into their existing or new Unity projects.
\paragraph{Import Splat} 
A splat PLY file is imported using the Splat Renderer block from our SplatXR Toolkit editor panel.
\paragraph{In-Editor Splat Rendering}
Developers can visualize and interact with the splat as a GameObject in the Unity viewport. Fig.~\ref{fig:lod_renderer} shows a visualization of the LoD rendering in a debug mode where Gaussians are colored based on the LoD they fall in.

\begin{figure}[H]
    \centering
    \includegraphics[width=0.88\linewidth]{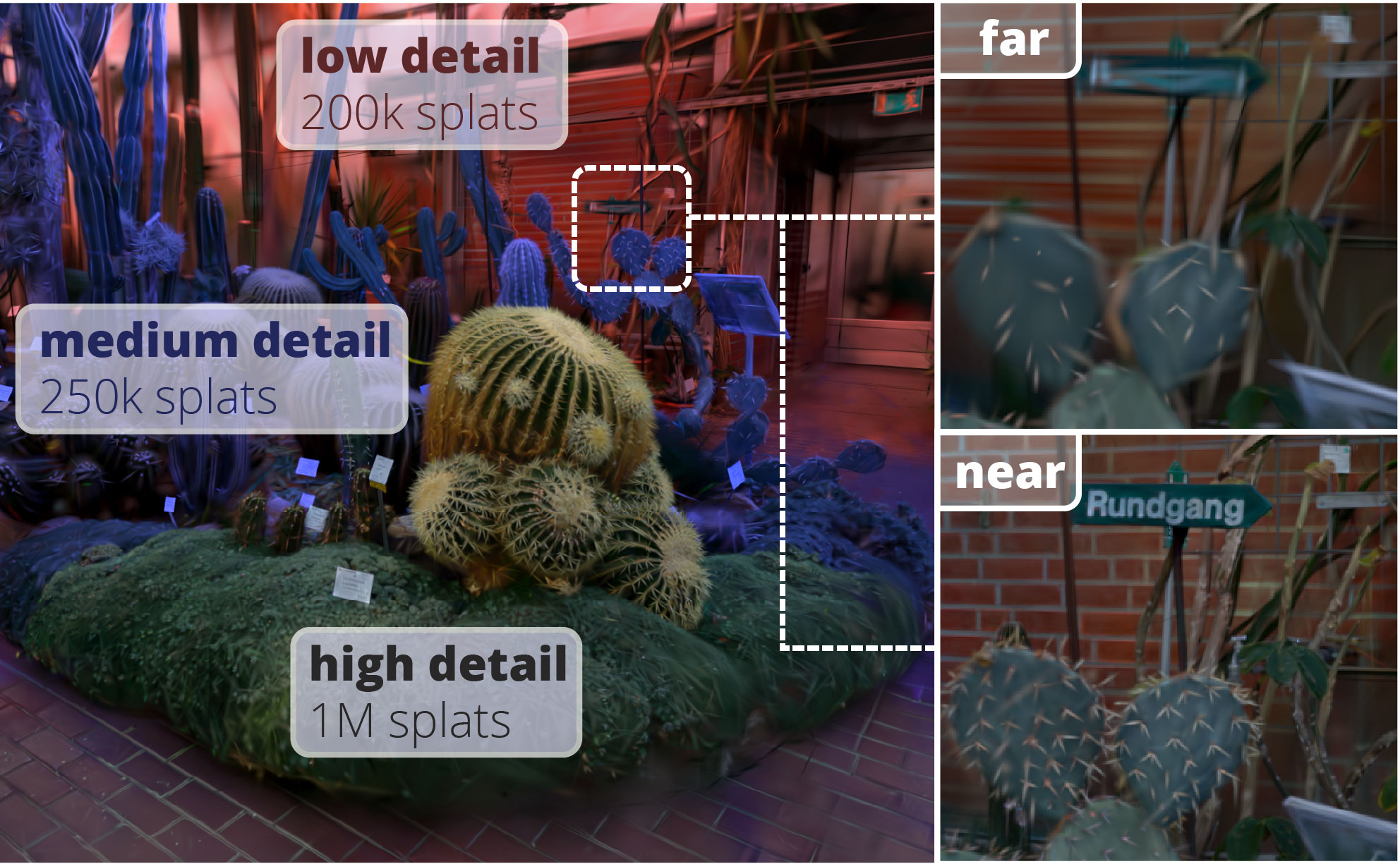}
    \Description{Figure 10 is composed of three subfigures. The first subfigure is larger than the two smaller subfigures to the right of it. The first subfigure shows the botanical garden. The closest part is showing a green cactus with yellow spines and a red-bricked floor in true colors. It is labelled "high detail, 1 million splats". The medium-range part, approx. two to five meters away from the observer is tinted in blue. It is labelled "medium detail, 250k splats", with a blue label. The far-range part, approx. beyond 5 meters away from the observer is tinted in red. It is labelled "low detail, 200k splats", with a red label. A small square is drawn onto a far away sign in the botanical garden. Two dashed lines lead to the small  subfigures, indicating that these subfigures zoom in on this sign. One small subfigure is captioned "far" showing the sign and a different cactus in a lot of blur such that the sign is illegible. The other small subfigure is captioned "near" showing the sign and the different cactus, very sharply.}
    \caption{\splatxr's level-of-detail rendering. \emph{Left:} Single LoD rendering. \emph{Right:} Comparison of the same volume rendered from far and near.}
    \label{fig:lod_renderer}
\end{figure}

\paragraph{SplatXR Object Structure}
Splats are created as GameObjects with associated scripts, shaders, and components in a hierarchical structure that separates the renderer from the empty parent object.
\paragraph{Splat Adjustments}
Splats' photometric properties can be adjusted from the shader panel.

\subsection{Technical Evaluation}
\label{app:technical-eval}
To evaluate our renderer, we displayed 9 splat scenes (shown in Fig.~\ref{fig:eval_scenes}) in a blank scene on Quest~3.
We have open sourced our evaluation setup on our project page.
To evaluate other toolkit functionality (meshing and splat import), we timed the operations in the Unity editor on a laptop PC (4070 GPU, 32 GB RAM).

\begin{figure}[H]
    \centering
    \includegraphics[width=0.54\linewidth]{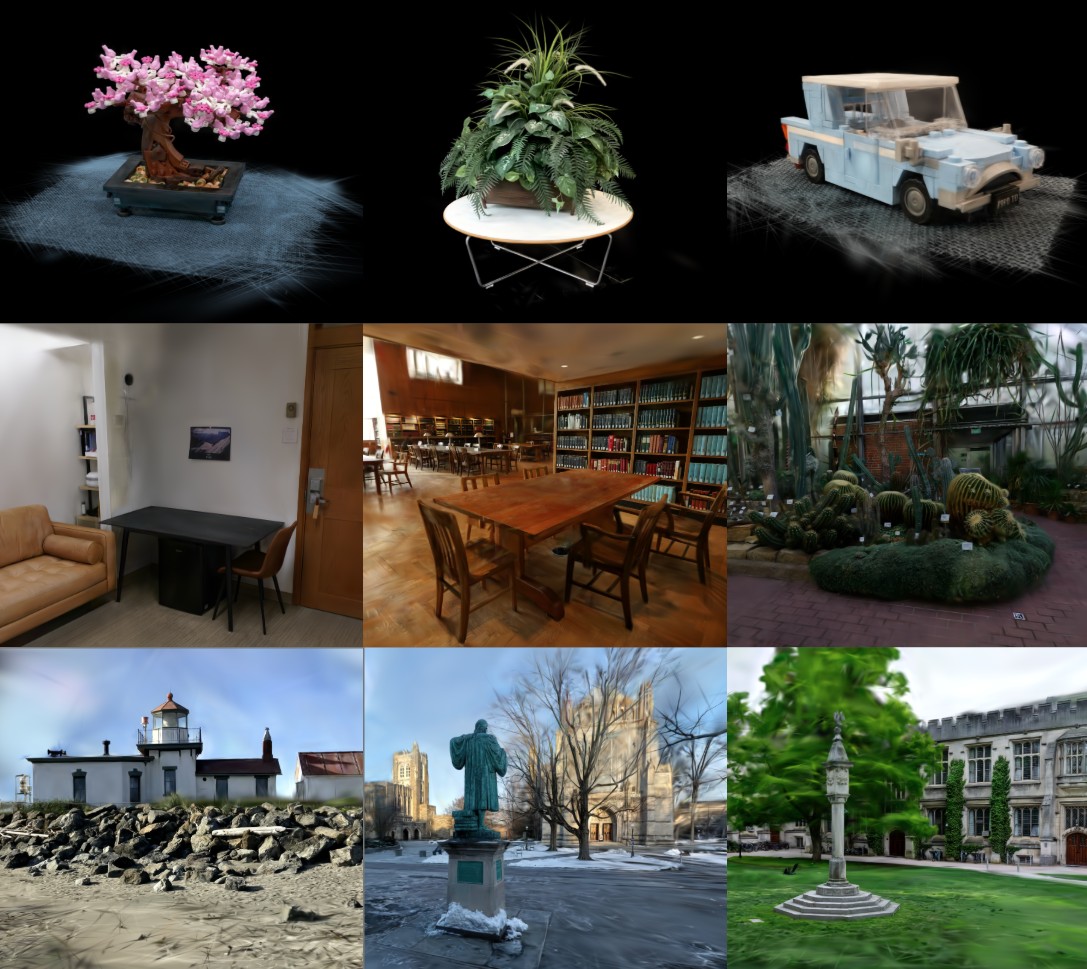}
    \Description{Figure 11 shows the 9 splat scenes used for the technical evaluation arranged in a 3x3 grid where the top row is the objects: a LEGO cherry blossom bonsai tree, a potted fern plant, and a LEGO blue car. The middle row shows indoor scenes: an office room, a library, and a botanical garden. The bottom row shows outdoor scenes: a lighthouse on a beach shore, an outdoor snowy courtyard, and a sundial in a field.}
    \caption{Splats used for technical evaluations: objects (top), indoor scenes (middle), and outdoor scenes (bottom).}
    \label{fig:eval_scenes}
\end{figure}

\begin{table}[H]
  \centering
  \caption{\revisionchanges{Renderer performance on a standalone Quest~3, averaged over three repetitions.}}
  \label{tab:renderer}
  \begin{tabular}{llcccc}
    \toprule
    & & \multicolumn{3}{c}{FPS} & GPU \\
    \cmidrule(lr){3-5}
    Scene type & Viewpoint & 50k & 100k & 200k & (ms) \\
    \midrule
    Object  & ---      & 72 & 72 & 59 & 7.6--15.4 \\
    \midrule
    Outdoor & Sparsest & 72 & 72 & 64 & 3--11  \\
            & Densest  & 72 & 53 & 38 & 12--26 \\
    \midrule
    Indoor  & Sparsest & 63 & 61 & 58 & ---    \\
            & Densest  & 58 & 46 & 36 & 13--32 \\
    \bottomrule
  \end{tabular}
\end{table}

\begin{table}[H]
  \centering
  \caption{\revisionchanges{Splat-to-mesh conversion time in seconds (mean $\pm$ SD) over the 27 splats from the rendering evaluation.}}
  \label{tab:mesh}
  \begin{tabular}{lccc}
    \toprule
    Algorithm & 50k & 100k & 200k \\
    \midrule
    Object (3 objects) & $25.7 \pm 3.9$ & $29.9 \pm 5.5$ & $41.4 \pm 4.3$ \\
    Space (6 scenes)   & $18.8 \pm 1.5$ & $22.7 \pm 1.1$ & $31.6 \pm 2.7$ \\
    \bottomrule
  \end{tabular}
\end{table}

}


\end{document}